\documentclass[fleqn,usenatbib]{mnras}

\usepackage{newtxtext,newtxmath}
\usepackage[T1]{fontenc}
\usepackage{ae,aecompl}

\usepackage{graphicx}	% Including figure files
\usepackage{amsmath}	% Advanced maths commands
\usepackage{amssymb}	% Extra maths symbols
\usepackage{subfigure}	
\usepackage{array}
\usepackage{booktabs} %set table
\usepackage{threeparttable}
\usepackage{textcomp,booktabs}
\usepackage[usenames,dvipsnames]{color}
\usepackage{colortbl}
\usepackage{soul}
\definecolor{mygray}{gray}{.9}
\definecolor{mypink}{gray}{.5}
\title[Spin of BH in 4U 1543-47]{The spin measurement of the black hole in 4U 1543-47 constrained with the X-ray reflected emission}

\author[Yanting Dong et al.]{
Yanting~Dong,$^{1,2}$\thanks{E-mail: ytdong@nao.cas.cn}
Javier~A.~Garc\'ia,$^{3,4}$
James~F.~Steiner,$^{5,6}$
Lijun~Gou$^{1,2}$\thanks{E-mail: lgou@nao.cas.cn}
\\
$^{1}$The National Astronomical Observatories, Chinese Academy of Sciences, Beijing, 100101, China\\
$^{2}$University of Chinese Academy of Sciences, No.19(A) Yuquan Road, Shijingshan District, Beijing, 100049, China\\
$^{3}$Cahill Centre for Astronomy and Astrophysics, California Institute of Technology, Pasadena, CA 91125, USA\\
$^{4}$Dr. Karl Remeis-Observatory and Erlangen Centre for Astroparticle Physics, Sternwartstr.~7, 96049 Bamberg, Germany
\\
$^{5}$CfA, 60 Garden St. Cambridge, MA 02138, USA\\
$^{6}$MIT Kavli Institute for Astrophysics and Space Research, MIT, 70 Vassar Street, Cambridge, MA 02139
}

\date{Accepted XXX. Received YYY; in original form ZZZ}

\pubyear{2018}

\begin{document}
\label{firstpage}
\pagerange{\pageref{firstpage}--\pageref{lastpage}}
\maketitle

\begin{abstract}
4U 1543-47 is a low mass X-ray binary which harbours a stellar-mass black hole located in our Milky Way galaxy. In this paper, we revisit 7 data sets which were in the Steep Power Law state of the 2002 outburst. The spectra were observed by the \emph{Rossi X-ray Timing Explorer}. We have carefully modelled the X-ray reflection spectra, and made a joint-fit to these spectra with {\tt relxill}, for the reflected emission. We found a moderate black hole spin, which is $0.67_{-0.08}^{+0.15}$ at 90\% statistical confidence. Negative and low spins (< 0.5) at more than 99\% statistical confidence are ruled out. In addition, our results indicate that the model requires a super-solar iron abundance: $5.05_{-0.26}^{+1.21}$, and the inclination angle of the inner disc is $36.3_{-3.4}^{+5.3}$ degrees. This inclination angle is appreciably larger than the binary orbital inclination angle ($\sim$21 degrees); this difference is possibly a systematic artefact of the artificially low-density employed in the reflection model for this X-ray binary system. 

\end{abstract}

\begin{keywords}
accretion, accretion discs -- X-rays: binaries -- stars: individual: IL Lupi, 4U 1543-47
\end{keywords}

%%%%%%%%%%%%%%%%%%%%%%%%%%%%%%%%%%%%%%%%%%%%%%%%%%

%%%%%%%%%%%%%%%%% BODY OF PAPER %%%%%%%%%%%%%%%%%%

\section{Introduction}
\label{sec:int}

An astronomical black hole can be readily characterized with two parameters, mass ($M$) and spin ($a_*$),  hence it could be described by Kerr metric \citep{ker1963}. Once we know these two parameters, we can make a complete description to the system. Compared to the mass, the spin is relatively harder to be constrained mainly because it only manifests in the most proximate, strong gravity region. The spin is commonly defined in terms of the dimensionless parameter $a_* = Jc/GM^2$ ($-1 \le a_* \le 1$, where $J$ is the angular momentum of the black hole, $c$ is the speed of light, and $G$ is the gravitational constant). 

As to the spin measurement, currently there are two leading approaches: the continuum-fitting method \citep{zha1997, Li2005} and the X-ray reflection fitting method \citep{iwa1997, mil2002}. Both approaches are based on the fundamental assumption that the inner edge of the accretion disc extends down to the innermost stable circular orbit ($R_\mathrm{in} = R_\mathrm{ISCO}$), which is a monotonic function of spin parameter $a_*$. $a_{*}=$ -1, 0 and 1 correspond to $R_\mathrm{ISCO}=$ 9, 6 and 1 $R_\mathrm{g}$, respectively, where $R_\mathrm{g}$ is the gravitational radius and is defined to be $R_\mathrm{g} = GM/c^2$. Given the inner radius of the accretion disc, one can readily obtain the spin.

The continuum-fitting method can measure the inner radius by modeling the thermal continuum of the accretion disc using {\tt kerrbb2}. {\tt kerrbb2} is a combination model of {\tt kerrbb} and {\tt bhspec}. {\tt kerrbb} has three fit parameters, $a_*$, the hardening factor $f$ and the mass accretion rate $\dot{M}$, only two of which can be determined at one time. A look-up table between $f$ and the scaled luminosity using {\tt bhspec} is generated. Then {\tt kerrbb} and the table allow one to directly fit for a* and $\dot{M}$ (refer sec. 4.2 of \citealt{mcc2006}). This method relies on accurate measurement to the system parameters of mass, distance, and inclination angle (often assumed to be identical to the orbital inclination angle) for the source \citep{gou2009, ste2011, che2016}. The X-ray reflection fitting method mainly models the relativistic reflection spectrum, which is a combination of fluorescent lines, absorption edges and recombination continua \citep{wan2017, wal2019, Gar2018}. One of the advantages of this technique is that it does not require information on the binary parameters, furthermore, it can make an independent constraint on the inclination angle of the inner disc. There have been consistent check for these two methods on several sources, and they generally showed consistent results \citep{rey2019}. 

It is expected that there exists billions of stellar-mass black holes in Milky Way galaxy \citep{bro1994,tim1996}, however, only roughly two dozen dynamically-confirmed black hole X-ray binaries have been confirmed \citep{rem2006}, and 4U~1543-47 (4U 1543) is one of them. This transient source was first discovered by \emph{Uhuru} satellite in 1971 \citep{mat1972}. Then, it went into outburst again in 1983, 1992 and 2002, respectively \citep{kit1984,har1992,par2004}. It was a very long time after the first discovery that the compact primary was confirmed to be a black hole \citep{rho1974,oro1998}. 

As to its spin parameter,  \citet{sha2006} first reported its spin with the continuum-fitting method. They estimated its spin to be $0.8\pm0.1$. Then, \citet{mil2009} and \citet{mor2014} reported two spin measurements, $0.3\pm0.1$ and $0.43^{+0.22}_{-0.31}$, respectively, both constrained by combining the continuum-fitting and the X-ray reflection fitting methods. These three works utilized the previous mass of $9.4 \pm{1.0} M_{\odot}$ and distance of $7.5 \pm{1.0}$ kpc, which were reported in \citet{par2004}. Except that \citet{mil2009} used the inclination angle of $32_{-4}^{+3}$ degrees constrained by their own fits to the iron line, the other two works used the inclination angle of $20.7 \pm{1.5}$ degrees which is equal to the binary orbital inclination angle. Recently, an updated set of dynamical parameters have been identified which have significant differences compared to earlier (J.Orosz, private communication).

In this paper, we revisited 7 \emph{Rossi X-ray Timing Explorer} (\emph{RXTE}, \citealt{zha1993}) data-sets of 4U 1543 to check the spin parameter via the X-ray reflection fitting method, i.e., carefully exploring the reflection component. We made a joint-fit for all spectra in order to achieve a better signal-to-noise ratio. we have adopted the updated reflection-emission model, {\tt relxill}~\citep{gar2014a, dau2014}. The whole paper is organized as follows. In Section \ref{sec:data}, we provide details of the data reduction and selection. In Section \ref{sec:results}, we describe the analysis of spectra and the spin result. In Section \ref{sec:disc} and Section \ref{sec:con}, we present our discussions and conclusions, respectively. 
%==============================================================

\section{Data Reduction and selection}
\label{sec:data}

\begin{figure}
\centering
	\includegraphics[scale=0.51,angle=0]{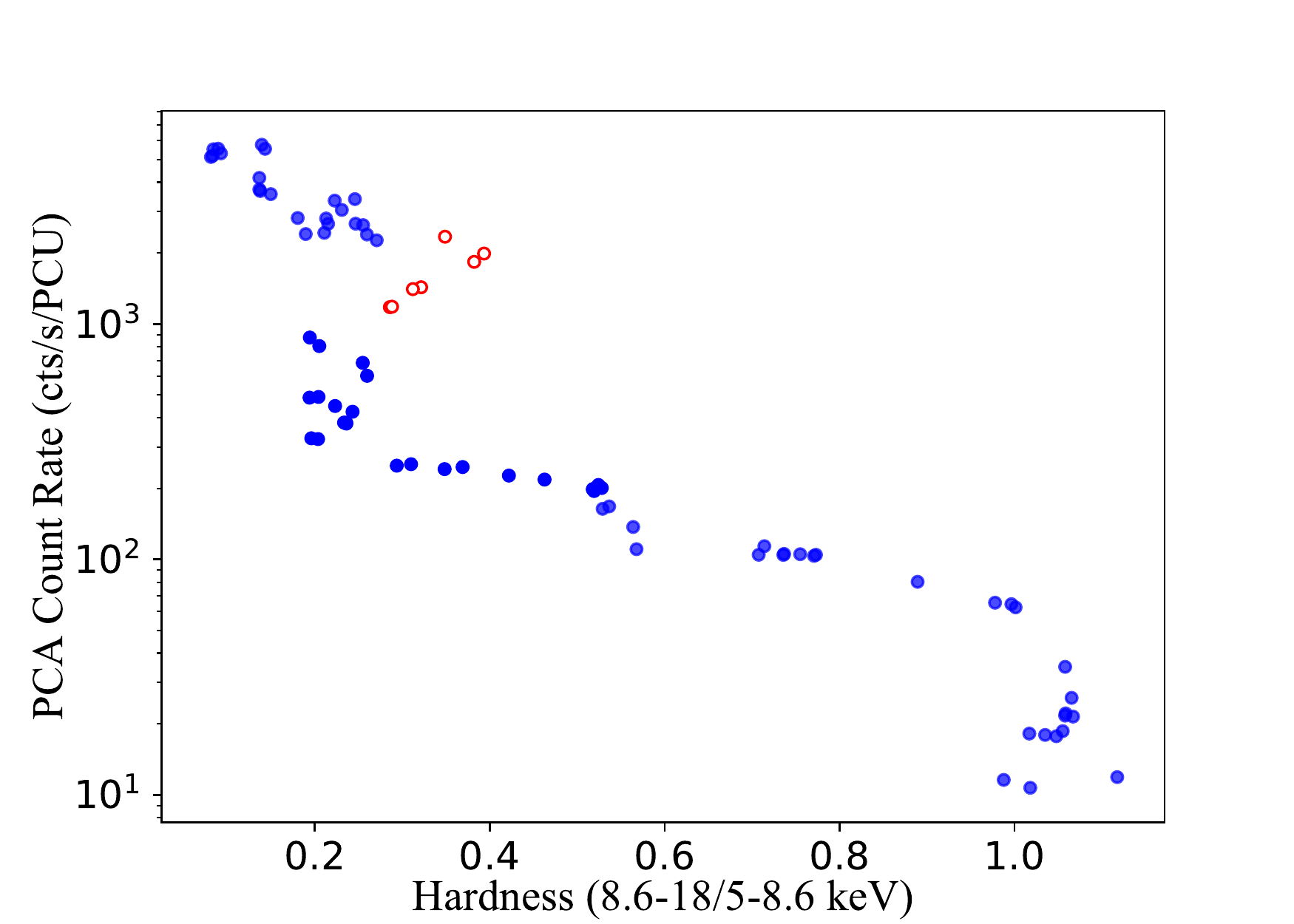}
    \caption{The evolutionary tracks in hardness-intensity diagram for observations except those with count rate smaller than 10 counts s$^{-1}$. The vertical axis presents the count rate in energy band 3-45 keV. The horizontal axis presents the hardness ratio (HR) defined as the ratio of count rate between 5-8.6 keV and 8.6-18 keV. %The observations whose Eddington-scaled unabsorbed luminosity span the range $\sim0.5-30$\% and outsiders are plotted as the open circles and solid circles, respectively.
    The 7 red open circles represent the data (two overlapping) we used to determine the spin of the black hole.}
    \label{fig:q}
\end{figure}

We revisited data sets for 4U~1543 which were observed by \emph{RXTE} during its 2002 outburst \citep{par2004}. There are 130 continuous pointed observations in total (the long exposures were split), collected by the Proportional Counter Array (PCA, \citealt{jah1996}). We only focused our analysis on the best-calibrated proportional counter unit, namely PCU2, as in previous work \citep{par2004, jah2006, sha2012}. All layers of the PCU2 were combined. We neglected 49 observations whose count rate is smaller than 10 counts s$^{-1}$. The remaining observations are presented in the hardness-intensity diagram (HID, Figure \ref{fig:q}).  
%==============================================================
%figre1
%==============================================================

\emph{RXTE}/PCA data of bright X-ray binaries are fundamentally limited not by counting statistics but by the systematic measure of calibration certainty in the detector. We apply a calibration correction, {\tt pcacorr} \citep{gar2014b}, which improves the instrumental response to a quality of 0.1\% precision. We include this 0.1\% as a systematic error. A second correction, {\tt crabcorr} \citep{ste2010}, standardizes the PCA absolute flux calibration to the \citet{too1974}  values for the Crab.  This latter tool improves not on the precision of the detector, but on the accuracy of our measurement.

We firstly subtracted background and made deadtime correction for the \emph{RXTE} data. Next, the calibration tool {\tt pcacorr} was applied. Then, a 0.1\% systematic error was added to the spectra. Finally, we performed \emph{RXTE} data analysis over the energy range between 2.8-45.0 keV using XSPEC 12.9.0g software package \citep{arn1996}. The quoted errors were given with a 90\% confidence level ($\Delta \chi ^2=2.71$)  if not specified. 

We selected 7 observations (MJD 52459-MJD 52463, defined as Spec. A-G) which show strong reflection components. In Table {\ref{tab:spec}}, we give the detailed information for these observations. In order to show the reflection features more clearly, we analysed the Spec. A-G between 2.8-45.0 keV, omitting 4.5-8.0 keV and 15.0-35.0 keV with the model {\tt crabcor*TBabs*(diskbb+powerlaw)} in XSPEC. 

For the model {\tt crabcor}, the normalization coefficient of $C=1.097$ and the slope difference of $\Delta\Gamma = 0.01$, are applied. For the model {\tt TBabs}, which is used to account for the galactic absorption by the interstellar medium (ISM) along the line of sight, the \citet{wil2000} set of solar abundances and the \citet{ver1996} photoelectric cross sections were specified accordingly. Since the effective low energy of \emph{RXTE} is limited at 2.8 keV, the data cannot constrain the column density ($N_\mathrm{H}$) well. The column density\footnote{\citet{mil2003} measured the column density for 4U 1543-47 to be (3.8 $\pm$ 0.2) $\times10^{21}$ cm$^{-2}$ by analysing the \emph{XMM-Newton}/EPIC-pn spectrum (0.3-10.0 keV). This value is consistent with the one we currently use in the fit. The new value of the column density has a negligible effect on our fitting results.} 
was fixed at $4.0\times10^{21}$ cm$^{-2}$ as in \citet{par2004} and \citet{mor2014}. 

\begin{figure*}
    \centering
	\includegraphics[scale=0.86,angle=0]{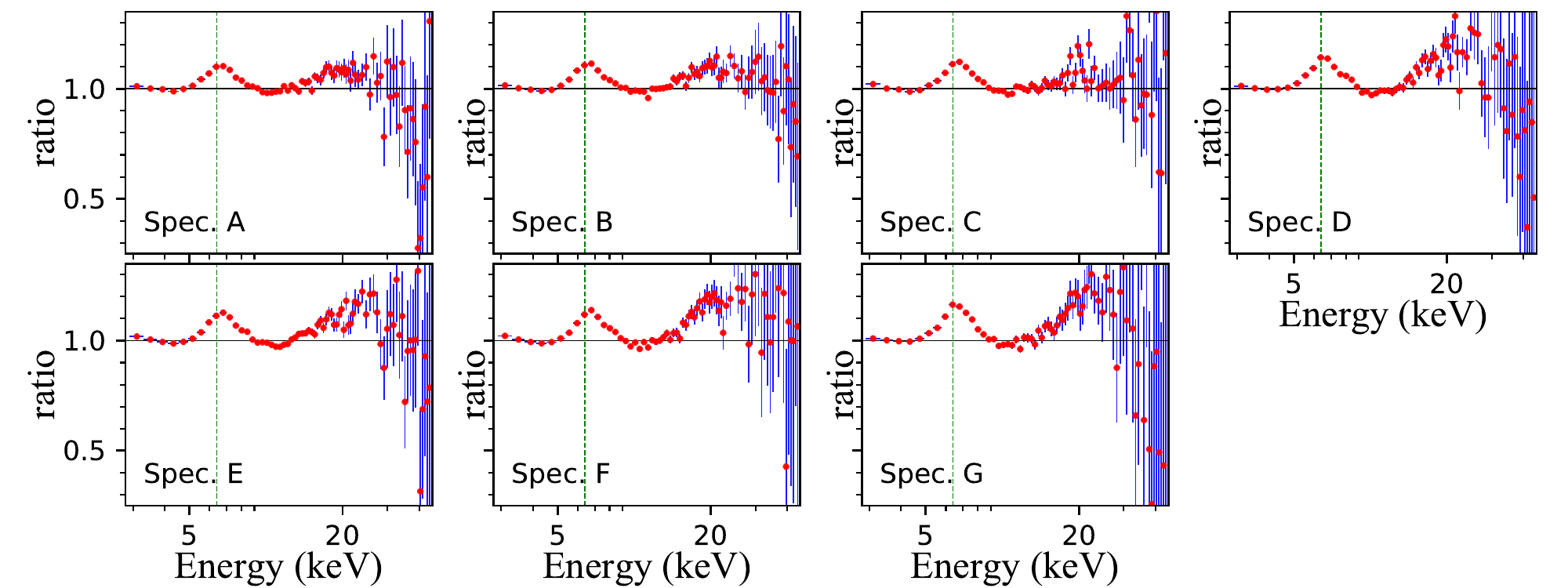}
    \caption{Spec. A-G were analysed using a simple mixture of an absorbed power-law together with multicolor disc blackbody model, respectively. The continuum models were fit over the energy band of 2.8-45.0 keV, ignoring 4.5-8.0 keV and 15-35keV region. Data-to-model ratios are plotted, with those regions added back in after the fit. The dotted green line represents the energy of 6.4 keV, which corresponds to the neutral iron line. The broad iron line and Compton hump, i.e. reflection component, are quite prominent.}
    \label{fig:ratio}
\end{figure*}

The fits to all 7 spectra are statistically unacceptable with $\chi^2_{\nu}$ = 31.32 (2129.78/68), 32.59 (2216.33/68), 23.26 (1581.93/68), 36.65 (2492.28/68), 63.75 (4335.57/68), 49.58 (3371.47/68) and 39.70 (2699.92/68), respectively. Data-to-model ratios are plotted in Figure {\ref{fig:ratio}}. The positive features in residuals are the broadened iron line and Compton hump characteristic of reflection emission.

%==============================================

%figre2
%==============================================================
%==============================================================

% TABLE 1
\begin{table*}
    \caption{Properties of Spec. A-G}
    \label{tab:spec}
    \newcommand{\tabincell}[2]{\begin{tabular}{@{}#1@{}}#2\end{tabular}}
    \begin{threeparttable}[b]
	\centering
	    \begin{center}
        \footnotesize
            \begin{tabular}{ccccccc}
            \toprule
            Spec.&Date&MJD&\tabincell{c}{Count Rate\\ (cts/s/PCU)} & \tabincell{c}{Exp.\\(s)}&\tabincell{c}{HR\\(8.6-18/5-8.6 keV)}&\tabincell{c}{$\chi^2_{\nu}$ $^a$\\(63 d.o.f)}\\
            \midrule
            A & July 04 & 52459 & 2348 & 1072 &0.35&67.70\\
            B & July 05 & 52460 & 1991  & 1120&0.39&63.76\\
            C & July 06 & 52461 & 1836 & 800  &0.38&77.15\\
            D & July 07 & 52462 & 1432 & 1328 &0.32&57.47\\
            E &  July 07 & 52462 & 1406 & 3376&0.31&63.49\\
            F & July 08  & 52463 & 1179  &3056&0.29&74.34\\
            G& July 08  & 52463 & 1184   &1520&0.29&83.88\\
            \bottomrule
            \end{tabular}
                \begin{tablenotes}
                \item[$a$]  The spectrum is fitted with {\tt crabcor*TBabs*smedge(diskbb+powerlaw+Gauss)} in XSPEC.
                 \end{tablenotes}
        \end{center}
    \end{threeparttable}
\end{table*}
%==========================================================

\section{Analysis and results}
\label{sec:results}

We fitted Spec. A-G with a phenomenological model, {\tt crabcor*TBabs*smedge(diskbb+powerlaw}\\
{\tt +Gauss)}, in which {\tt Gauss} and {\tt smedge} \citep{ebi1994} are used to model the reflection features. The central energy of the iron line was constrained between 6.0 and 6.97 keV. The width and the normalization were allowed to be free. The width of {\tt smedge} is fixed at 7.0 keV, the smeared edge could change from 7.0 to 9.0 keV, and the optical depth floated freely. We only focused on the best-fitting models. The detailed information on the quality of the fit for each spectrum is shown in Table \ref{tab:spec}. The temperatures of thermal emission are $\sim$0.75-0.85 keV. The photon indexes of power-law emission are $\sim$2.45-2.75, which indicate that the source is in the Steep Power Law (SPL) state during these observations. Line peaks are less than 6.4 keV, which suggest the presence of strong gravitational redshift around the black hole. 

We then fit the full ionized reflection spectrum with an sophisticated model. The reflection information, however, was weak relative to the disc and corona continua, we, therefore, make a joint-fit to Spec. A-G. We used a relativistic reflection model {\tt relxill} \citep{gar2014a, dau2014} which is a combination of the reflection model {\tt xillver} \citep{gar2010, gar2011, gar2013} and the relativistic convolution kernel {\tt relconv} \citep{dau2010, dau2012, dau2013}. This model is designed to fit the reflection and the power-law components simultaneously. It has been widely used in recent years for the reflection exploration in stellar-mass black hole binaries and AGNs, sometimes also in neutron star binaries. The returned parameter list contains inner index ($q_\mathrm{in}$), outer index ($q_\mathrm{out}$), and break radius ($R_\mathrm{br}$) which describe the radial dependence of the emissivity of reflection emission; spin parameter ($a_*$), inclination angle ($i$), inner radius ($R_\mathrm{in}$), outer radius ($R_\mathrm{out}$), redshift ($z$) to the source (set to 0 for Galactic systems), photon index ($\Gamma_\mathrm{r}$), ionization state ($\mathrm{log}\xi$), iron abundance ($A_\mathrm{Fe}$), high energy cut-off ($E_\mathrm{cut}$), reflection fraction ($R_\mathrm{f}$), and normalization ($N_\mathrm{r}$).

%%%%% today

The overall self-consistent model we adopt here is {\tt crabcor*TBabs(diskbb+relxill)}. For {\tt relxill} model, we assumed a single emissivity profile ($q_\mathrm{in}=q_\mathrm{out}=q$) and the inner radius of accretion disc extended down to the ISCO radius ($R_\mathrm{in} = R_\mathrm{ISCO}$). Some parameters were independent for each spectrum: the temperature $T_\mathrm{col}$ and normalization constant $N_\mathrm{DISC}$ of thermal emission; the emissivity index $q$, photon index $\Gamma _\mathrm{r}$, ionization state $\mathrm{log}\xi$, reflection fraction $R_\mathrm{f}$ and normalization $N_\mathrm{r}$ of reflection component. The other parameters were linked together among 7 spectra. The spin parameter $a_*$ and the inclination angle $i$ were free. The outer radius was set to default value: $R_\mathrm{out}=400$ $R_\mathrm{g}$. Because the power-law is extremely steep, we can't detect the high energy cut-off in these observations. We, then, fixed $E_\mathrm{cut}$ at 300~keV which is a physically reasonable and sufficiently large value for our purposes. Meanwhile, it is beneficial for reducing the complexity of the model. When the iron abundance was fixed at unity (i.e. solar abundance), the model returned an acceptable but not a good fit with $\chi^2_{\nu}$ = 1.56 (707.96/453), and the spin tended to peg at the maximal negative value of -0.998. Therefore, we let the iron abundance $A_\mathrm{Fe}$ free.

%==============================================================
%Table 2
\begin{table*}
    \caption{Best-fitting parameters for Model 1: {\tt crabcor*TBabs}({\tt diskbb}+{\tt relxill})}
    \begin{threeparttable}[b]
    \begin{center}
    \label{tab:relxill}
    \footnotesize
        \begin{tabular}{llccccccc}
        \toprule
        Model & Parameter &Spec. A & Spec. B & Spec. C & Spec. D & Spec. E & Spec. F & Spec. G \\
        \midrule
        {\tt crabcor}&$C$  &\multicolumn{7}{c}{1.097  (f)}\\
        \specialrule{0em}{1pt}{1.1pt}
        &$\Delta\Gamma$  &\multicolumn{7}{c}{0.01  (f)}\\
        \specialrule{0em}{1pt}{1.1pt}
        {\tt TBabs}&$N_\mathrm{H}$ ( cm$^{-2}$) &\multicolumn{7}{c}{$4.0\times10^{21}$  (f)}\\
        \specialrule{0em}{1pt}{1.1pt}
        {\tt diskbb}    &$T_\mathrm{col}$ (keV)&$0.858_{-0.014}^{+0.011}$&$0.825\pm0.012$&$0.835_{-0.018}^{+0.015}$&$0.787_{-0.010}^{+0.008}$&$0.787_{-0.009}^{+0.007}$&$0.771\pm0.007$&$0.772_{-0.009}^{+0.010}$\\
        \specialrule{0em}{1pt}{1.1pt}
        &$N_\mathrm{DISC}$&$3889_{-219}^{+366}$&$4151_{-273}^{+237}$&$3561_{-297}^{+312}$&$4862_{-247}^{+329}$&$4923_{-234}^{+398}$&$5025_{-212}^{+230}$&$4891_{-311}^{+328}$\\
        \specialrule{0em}{1pt}{1.1pt}
        {\tt relxill}
        &$q$&$3.69_{-0.56}^{+1.03}$&$3.78_{-0.64}^{+1.23}$&$3.29_{-0.52}^{+1.04}$&$3.98_{-0.70}^{+1.41}$&$3.82_{-0.57}^{+1.09}$&$3.67_{-0.55}^{+1.20}$&$3.71_{-0.65}^{+1.13}$ \\
        \specialrule{0em}{1pt}{1.1pt}
        &$a_*$&\multicolumn{7}{c}{$0.67_{-0.08}^{+0.15}$}\\
        \specialrule{0em}{1pt}{1.1pt}
        &$i$ (deg)&\multicolumn{7}{c}{$36.3_{-3.4}^{+5.3}$ }\\
        \specialrule{0em}{1pt}{1.1pt}
        &$A_\mathrm{Fe}$&\multicolumn{7}{c}{$5.05_{-0.26}^{+1.21}$}\\
        \specialrule{0em}{1pt}{1.1pt}
        &$\Gamma_\mathrm{r}$&$2.79\pm0.06$&$2.59_{-0.04}^{+0.06}$&$2.65_{-0.07}^{+0.08}$&$2.71_{-0.09}^{+0.06}$&$2.66_{-0.08}^{+0.07}$&$2.69_{-0.08}^{+0.06}$&$2.80_{-0.07}^{+0.12}$ \\
        \specialrule{0em}{1pt}{1.1pt}
        &$\mathrm{log}\xi$&$3.72_{-0.16}^{+0.13}$&$3.54_{-0.13}^{+0.22}$&$3.53_{-0.15}^{+0.46}$&$3.70_{-0.21}^{+0.13}$&$3.72_{-0.19}^{+0.29}$&$3.73_{-0.12}^{+0.15}$&$3.71_{-0.18}^{+0.05}$\\
        \specialrule{0em}{1pt}{1.1pt}
        &$R_\mathrm{f}$&$0.35_{-0.10}^{+0.07}$&$0.29_{-0.03}^{+0.06}$&$0.27_{-0.04}^{+0.07}$&$0.44_{-0.11}^{+0.05}$&$0.41_{-0.08}^{+0.06}$&$0.44_{-0.07}^{+0.16}$&$0.61_{-0.14}^{+0.12}$  \\
        \specialrule{0em}{1pt}{1.1pt}
        &$N_\mathrm{r}$&$0.184_{-0.031}^{+0.038}$&$0.091_{-0.011}^{+0.016}$&$0.099_{-0.022}^{+0.024}$&$0.065_{-0.016}^{+0.013}$&$0.051_{-0.012}^{+0.011}$&$0.04_{-0.009}^{+0.008}$&$0.057_{-0.012}^{+0.026}$   \\
        \midrule
        & $\chi^2$ &\multicolumn{7}{c}{390.40} \\
        & $\nu$ & \multicolumn{7}{c}{452} \\
        & $\chi^2_{\nu}$ & \multicolumn{7}{c}{0.86} \\
        \bottomrule
        \end{tabular}
    \begin{tablenotes}
     \item \textbf{Notes.} Columns 3-9 show successively the results of Spec. A-G. The parameters with ``f'' in parenthesis indicate they were fixed at values given. All errors for one parameter of interest were calculated with 90\% confidence level.
     \item \textbf{1.} Parameters including the spin $a_*$, the inclination angle $i$ and the iron abundance $A_\mathrm{Fe}$ of {\tt relxill} were linked together among different spectra.
     \item \textbf{2.} Parameters including the temperature $T_\mathrm{col}$, normalization constant $N_\mathrm{DISC}$, emissivity index $q$, photon index $\Gamma _\mathrm{r}$, ionization state $\mathrm{log}\xi$, reflection fraction $R_\mathrm{f}$ and the normalization $N_\mathrm{r}$ were independent for each spectrum.
    \end{tablenotes}
    \end{center}
    \end{threeparttable}
\end{table*}
%==============================================

%==============================================

%figre3
\begin{figure*}
    \centering
	\includegraphics[scale=0.50,angle=0]{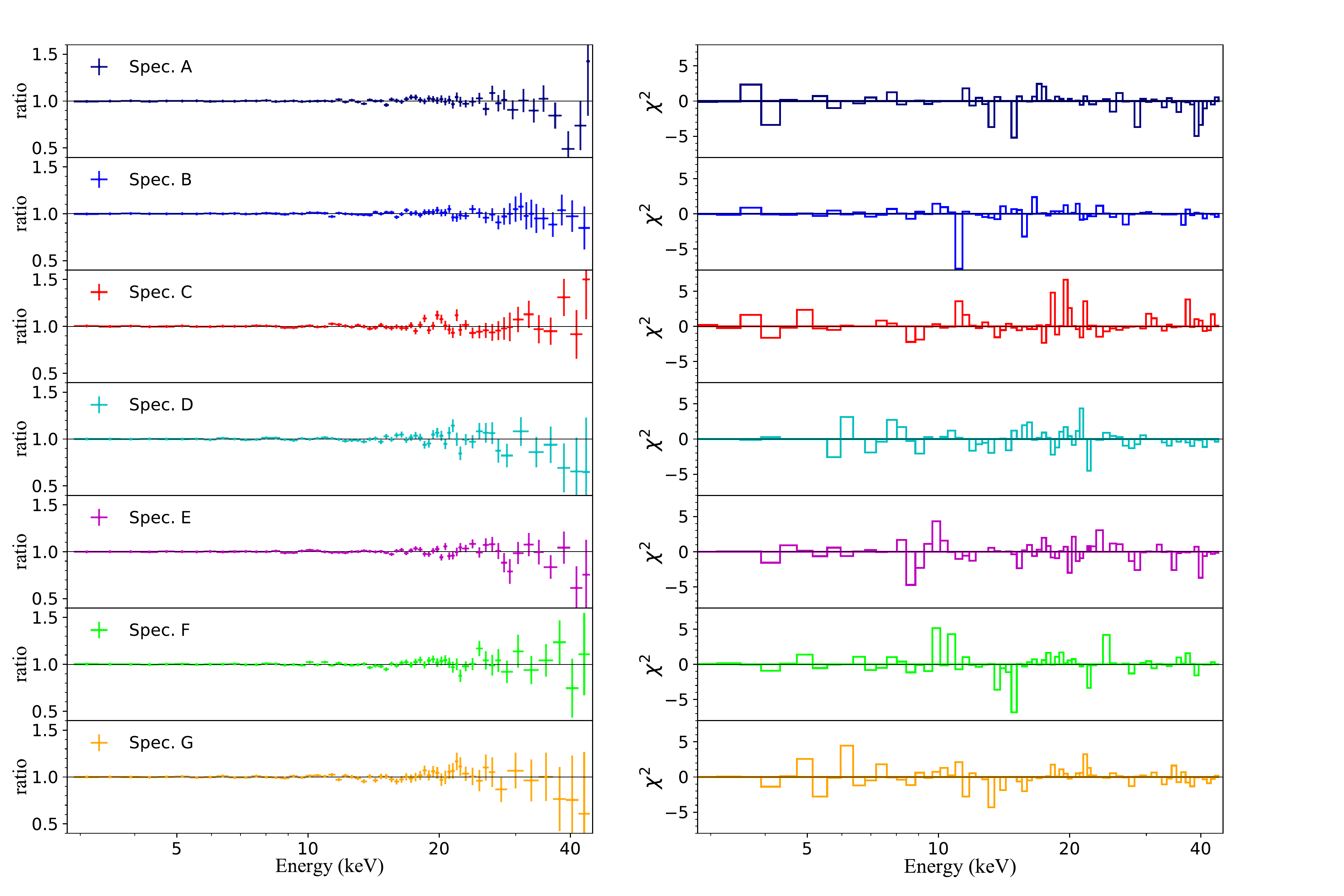}
    \caption{Making joint-fit to Spec. A-G using model: {\tt crabcor*TBabs*(diskbb+relxill}). Data-to-model ratios and contributions to the total $\chi^2$ are presented in left and right panels, respectively. Different colors represent different spectra. The model achieved a satisfactory fit with $\chi^2_{\nu}=390.40/452$.}
    \label{fig:relxill}
\end{figure*}

%==============================================================
%==============================================================
The model achieved a statistically good fit with $\chi^2_{\nu}$ = 0.86 (390.4/452) for 7 observations (Table \ref{tab:relxill}). All parameters including the spin and the inclination angle are well constrained. The spin parameter $a_*$ is obtained to be $0.67_{-0.08}^{+0.15}$. The inclination angle $i$ is obtained to be $36.3_{-3.4}^{+5.3}$ degrees. The iron abundance $A_\mathrm{Fe}$ is obtained to be $5.05_{-0.26}^{+1.21}$. Figure \ref{fig:relxill} shows the data-to-model ratios and the contributions to the total $\chi^2$ of the best-fitting. No distant reflection from the outer disc, the wind or the surface of companion \citep{wan2018, xu2018}, was necessary, which is attributed to that the \emph{RXTE} is not sensitive to the narrow line. When the distant reflection component is added using {\tt xillver}, the statistic is not improved with $\chi^2_{\nu}$ = 0.87 (385.82/445). The spin is $0.78_{-0.13}^{+0.98}$, and the inclination angle is $41.35_{-6.69}^{+10.86}$ degrees, which is still consistent with the model without {\tt xillver}.

In order to investigate the effect of different values of column density on our model, especially the main parameters, we tried to let the parameter $N_\mathrm{H}$ free. However, the model was unable to provide any meaningful constraint on $N_\mathrm{H}$ (a detection of $N_\mathrm{H}$ only 1$\sigma$). This is not surprising given the band of sensitivity for the PCA. Most importantly, this has negligible impact on the fit parameters for the model.

We investigated the model dependence on the high energy cut-off. We made $E_\mathrm{cut}$ vary among 7 observations. Comparing with the best-fitting result we describe above, the fit was only improved with $\Delta \chi^2 = 9.33$ for reducing 7 degrees of freedom (d.o.f), which is not a significant improvement. Meanwhile, the fit did not constrain $E_\mathrm{cut}$ well. This changed setting for $E_\mathrm{cut}$ did not affect profiles of thermal, power-law and reflection emission largely. It still requires a higher super-solar iron abundance of 6.61$_{-1.17}^{+2.06}$. Moreover, the free $E_\mathrm{cut}$ did not change the inclination angle and the spin of the black hole largely. The spin parameter $a_*$ is obtained to be $0.73_{-0.14}^{+0.10}$. The inclination angle $i$ is obtained to be $34.2_{-3.7}^{+6.8}$ degrees.

For the broken emissivity profile, i.e. the inner index is free, the outer index is fixed at 3, and the break radius is fixed at 15 $R_{\rm{g}}$, is assumed, the fit statistics are not improved with $\chi^2_{\nu} = 0.86$ (388.72/452). It is also found that the best-fitting parameters are not changed significantly, such as the spin parameter $a_*$ is $0.71_{-0.08}^{+0.11}$, the inclination angle $i$ is $35.8_{-2.7}^{+3.6}$ degrees, and the inner indexes are $\sim3-4$. For the single emissivity profile, we explored the implication of freezing the emissivity indexes at typical value ($q$ = 3) instead of being free. The best-fitting model only gave a lower limit 0.65 of spin at 90\% confidence level with $\chi^2_{\nu} = 0.88$ (403.92/459), which is slightly worse than the case with free $q$ ($\Delta\chi^2=13.52$ for increasing 7 d.o.f). The inclination angle $i$ is also greater than the orbital inclination angle by approximately 10$^{\circ}$. 

\begin{figure}
    \centering
	\includegraphics[scale=0.51,angle=0]{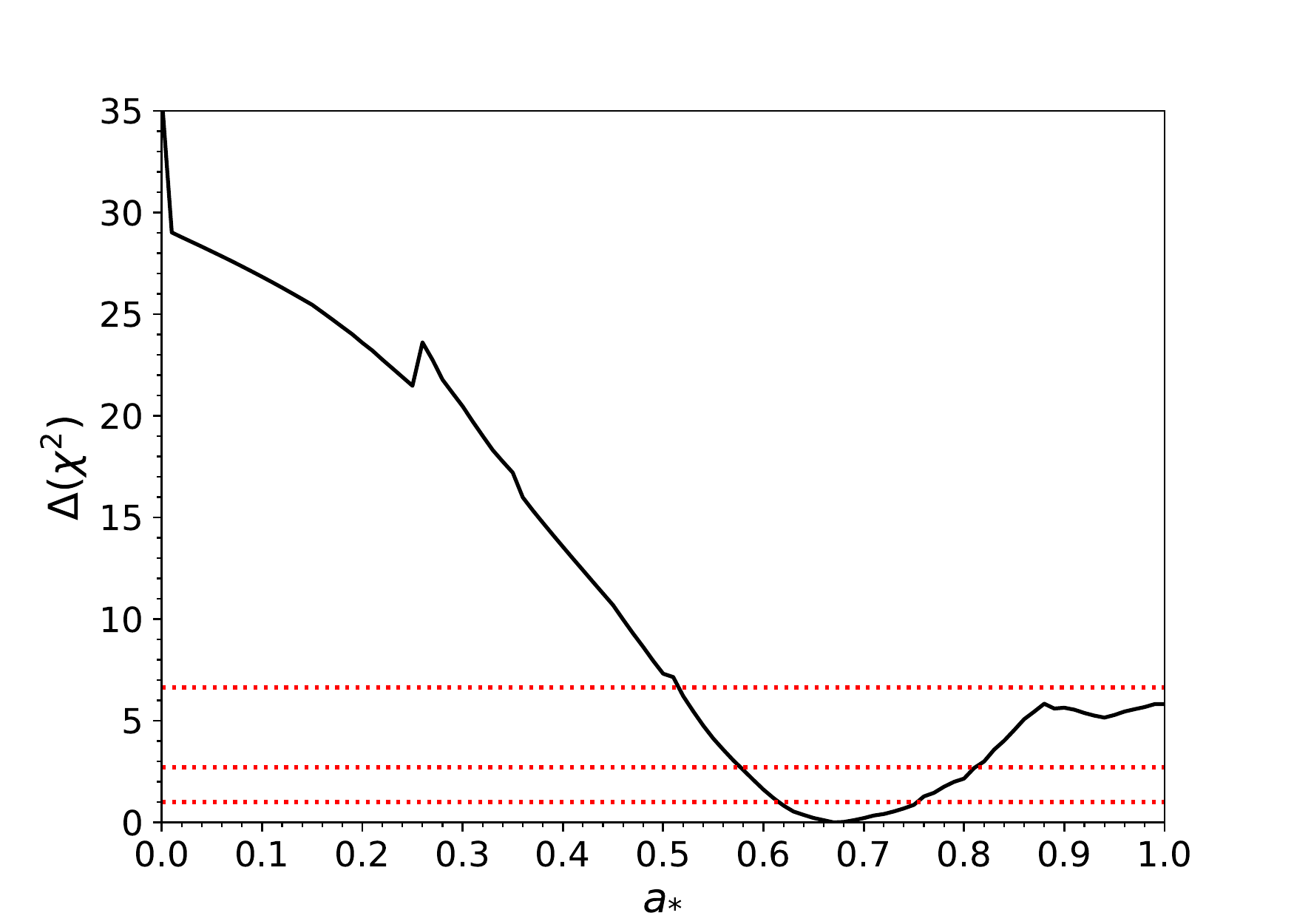}
    \caption{Joint-fit to Spec. A-G using model {\tt crabcor*TBabs*(diskbb+relxill)} (Model 1). The goodness-of-fit statistic as a function of the black hole spin parameter $a_{*}$ is shown in the above plot. The stepsize of 0.01 was explored from 0.0 to 1.0. Confidence levels of 68\%, 90\%, and 99\% are labeled with red dotted lines. It suggests a moderate rotating black hole and strongly excluded negative and low spins.}
    \label{fig:a}
\end{figure}

\begin{figure}
    \centering
	\includegraphics[scale=0.51,angle=0]{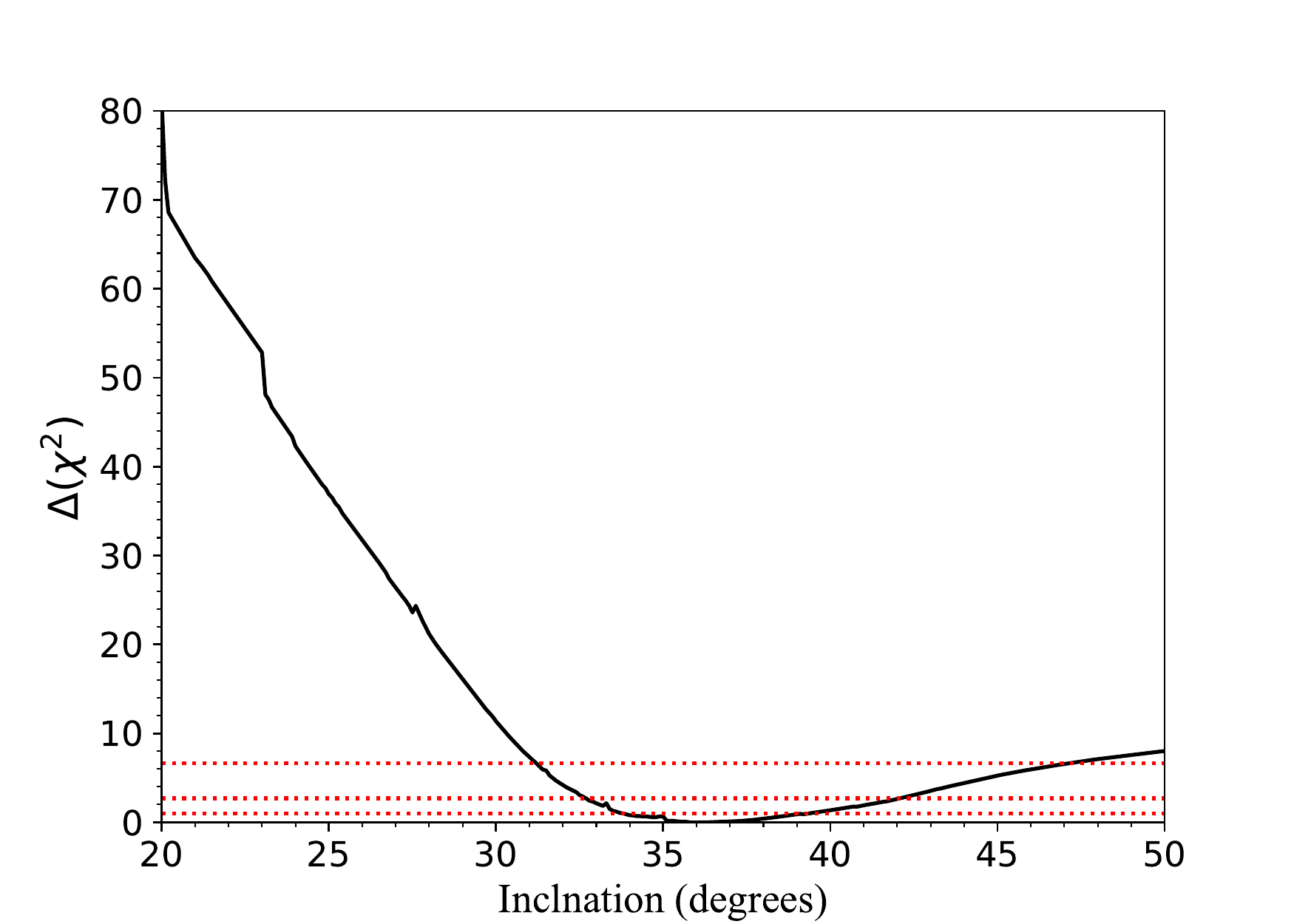}
	\caption{Joint-fit to Spec. A-G using model {\tt crabcor*TBabs*(diskbb+relxill}) (Model 1). The goodness-of-fit statistic as a function of the accretion disc inclination angle parameter $i$ is shown in the above plot. The stepsize of 0.1$^{\circ}$ was explored from 20$^{\circ}$ to 50$^{\circ}$. Confidence levels of 68\%, 90\%, and 99\% are labeled with red dotted lines. It indicates the inclination angle is larger than $\sim$32$^{\circ}$ and smaller than $\sim$40$^{\circ}$ at 90\% confidence.}
    \label{fig:i}
\end{figure}

We explored the $\chi^2$ parameter space using the command ``steppar'' for the spin and the inclination angle. During the searching process, at each step the parameters of interest were fixed at incrementally stepped values while all other parameters were allowed to fit. For the spin parameter, the stepsize of 0.01 was used from 0.0 to 1.0 (Figure \ref{fig:a}). For the inclination angle parameter, the stepsize of 0.1$^{\circ}$ was explored from 20$^{\circ}$ to 50$^{\circ}$ (Figure \ref{fig:i}). Three levels of confidence (68\%, 90\% and 99\%) are also marked in both figures. The spin is constrained well between $\sim{0.58-0.82}$ at 90\% statistical confidence which is consistent with a moderate spin black hole. Negative and low spins (< 0.5) at more than 99\% statistical confidence are ruled out. The inclination angle is constrained to be $\sim{32^{\circ}-42^{\circ}}$ at 90\% statistical confidence.
%==============================================================

%figre4

%==============================================================
%figure5
%==============================================

%==============================================
\section{Discussions}
\label{sec:disc}
In this paper, we have carefully explored the constraint on the spin of the black hole in 4U 1543 on the basis of its reflection emission. We selected 7 SPL state spectra which show strong reflection component. These spectra were selected from the 2002 outburst observed by \emph{RXTE}.

According to the phenomenological model of {\tt crabcor*TBabs*smedge(diskbb+powerlaw+\\Gauss)} in Section \ref{sec:results}, the central energy of the Gaussian profile is less than 6.4 keV, suggesting the presence of strong gravitational redshift around the black hole, and the reflection region is concentrated quite close to the black hole. To improve sensitivity to faint reflection features, we have fitted the 7 spectra simultaneously \citep{gar2015}. \citet{mil2009} found that {\tt diskbb} and physically rigorous relativistic disc models such as {\tt kerrbb} performed similarly in their ability to characterize the thermal continuum for the purposes of isolating the reflection signal.

From the best-fitting parameters in Table \ref{tab:relxill}, the color temperature $T_\mathrm{col}$ of accretion disc drops down with decreasing luminosity. The relative consistency of $N_\mathrm{DISC}$ indicates that the disc radius and coronal covering cannot change appreciably over the 5-day span in which these observations were accrued. The photon index is relatively constant and constrained between 2.6-2.8. The emissivity index is also relatively constant but larger than the canonical value ($q=3$). The high values of $\mathrm{log}\xi$ (> 3.5) indicate that the accretion disc is highly ionized. The reflection fraction $R_\mathrm{f}$, which is defined as the ratio that the power-law emission hit on the disc to that escaped to the infinite, is in the range of $\sim{0.25-0.65}$. It goes up modestly with decreasing luminosity. Because the hot gas layer at the surface of the disc dilutes the reflection signal by scattering and blurring reflection features \citep{nay2001}. The hotter the surface of disc, the more significant the dilution is.

\begin{figure}
\centering
	\includegraphics[scale=0.51,angle=0]{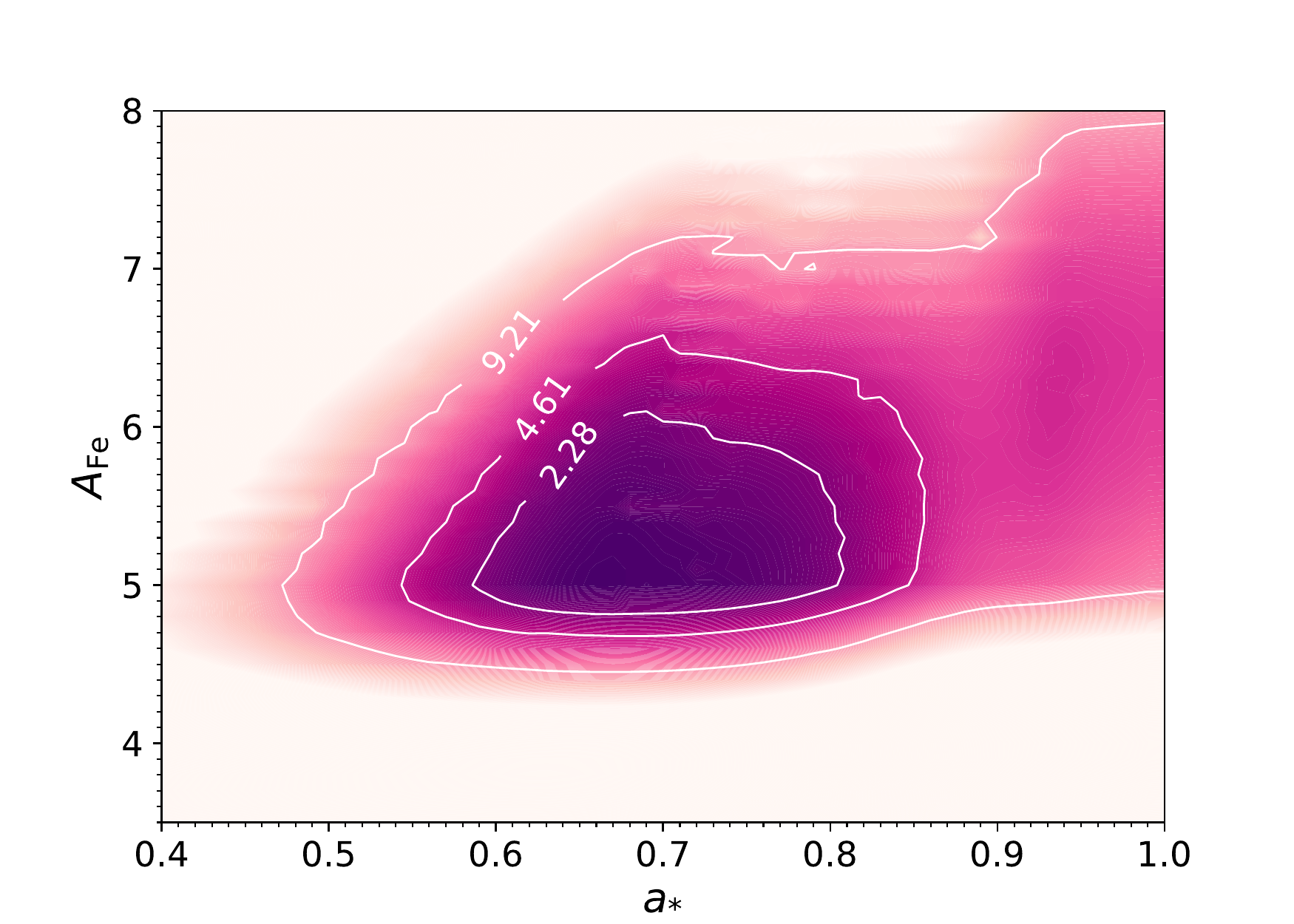}
    \caption{Joint-fit to Spec. A-G using model {\tt crabcor*TBabs*(diskbb+relxill)} (Model 1). Contours of $\chi^2$ with 68\%, 90\% and 99\% confidence levels for the spin parameter $a_{*}$ and iron abundance parameter $A_\mathrm{Fe}$ were measured using reflected components and shown in the above plot. By dividing the spin between 0.4-1.0 into 60 evenly spaced values and fixing the spin at certain value each time, we then fit spectra for the iron abundance, and it is found that 50 out of 60 fixed spin values give the $A_\mathrm{Fe}$ in the range of $3.5-8.0$. It shows a weakly positive relationship between them.}
    \label{fig:relxill_aFe}
\end{figure}

We also report a super-solar iron abundance for the binary system 4U 1543. Based on the full reflection Model l, we investigated the degeneracy between the iron abundance and the spin by fixing the spin value and fitting for the iron abundance. We divided the spin between the range of 0.4-1.0 into 60 evenly-spaced values, and it is found that 50 of 60 fitted iron abundance $A_{Fe}$ is in the range of 3.5-8.0 in units of solar abundance (Figure \ref{fig:relxill_aFe}). The relation indicates that the iron abundance has a weakly positive correlation with the spin. The iron abundance is greater than 4 given the 99\% confidence range level. 

The very large iron abundance is not unique to 4U 1543. Similar results have been reported in other stellar-mass black hole binaries such as GX 339-4 ($A_\mathrm{Fe}=5\pm1$ solar in \citealt{gar2015} and $A_\mathrm{Fe}=6.6\pm0.5$ solar in \citealt{par2016}), V404 Cyg ($A_\mathrm{Fe}\sim{5}$ solar in \citealt{wal2017}), and Cyg X-1 ($A_\mathrm{Fe}=4.7\pm0.1$ solar in \citealt{par2015} and $A_\mathrm{Fe}=$4.0-4.3 solar in \citealt{wal2016}). At present, there is no satisfactory physical explanation for the occurrence of high iron abundance in these systems. The most likely explanation is the atomic data shortcomings in current reflection models. \citet{joh2018} explored the super-solar iron abundance of Cyg X-1 using observation of the intermediate state. They found that the higher electron density ($n_{e} \approx 4 \times 10^{20}$ cm$^{-3}$) model was compatible with solar iron abundance using the high-density model {\tt reflionx\_hd} (a new version of {\tt reflionx}). However, the range of the photon index in that model is 1.4-2.3, which can not be applied to the observations in our paper. While the maximum density in high-density version of {\tt relxill} ({\tt relxillD}) is only 10$^{19}$ cm$^{-3}$. When it is used to fit the data, the density is pegged to its upper limit, which indicates that the density in the disc is larger than the maximal value in the model.

\begin{figure}
\centering
	\includegraphics[scale=0.51,angle=0]{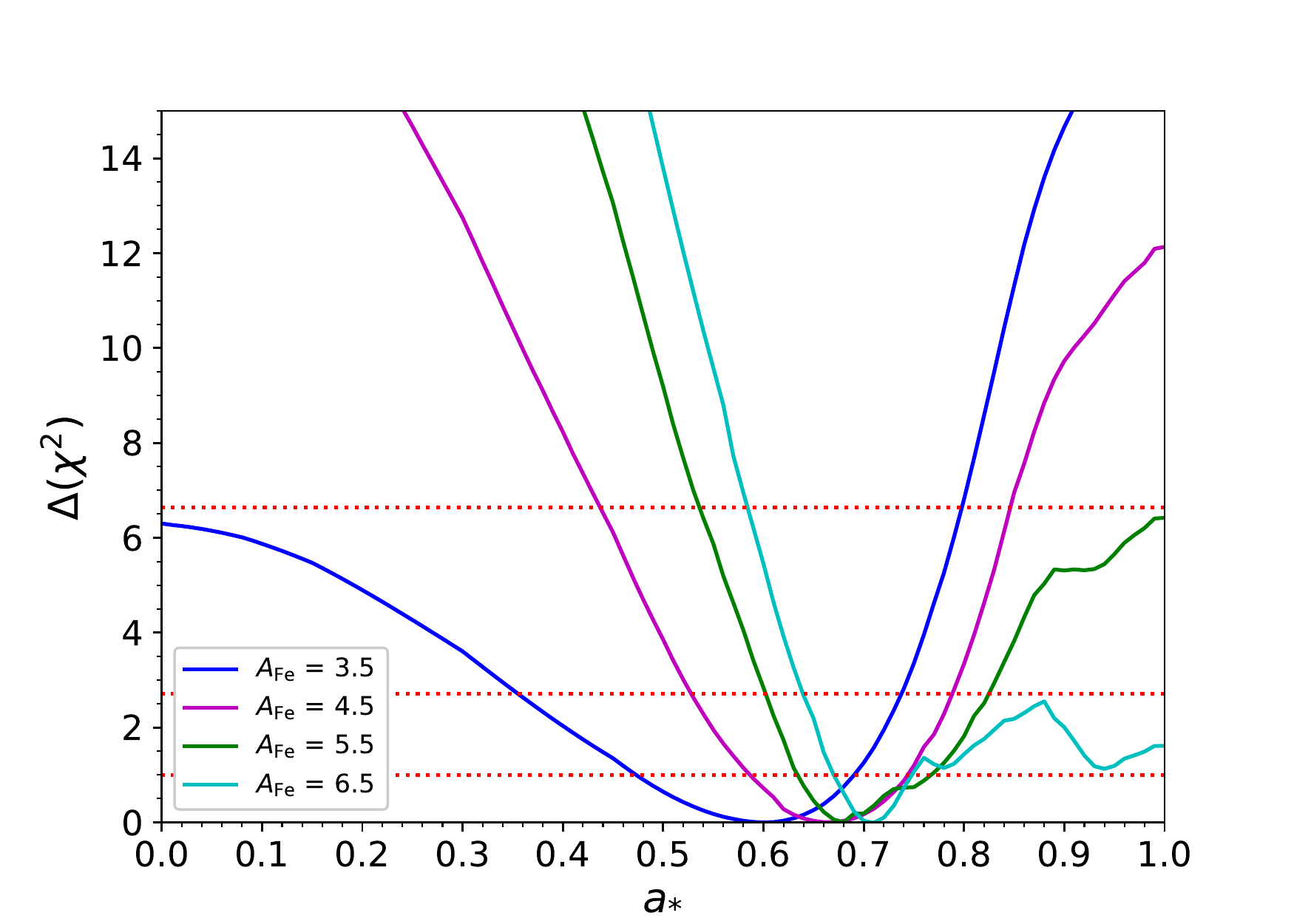}
    \caption{Fixing the iron abundance from 3.5 to 6.5 with a stepsize of 1.0, respectively. The 4 contours for $\chi ^2$ dependence on the spin with 68\%, 90\% ,and 99\% confidence levels are shown.}
   \label{fig:Fe_step}
\end{figure}

We found that the spin parameter pegged at -0.998 when the iron abundance was decreased to $\sim{3}$. To better understand this surprising finding, we did another trial. We fixed the iron abundance at values between 3.5-6.5 with a stepsize of 1.0, and found four best fits with $\chi^2_{\nu}$ < 1 (d.o.f = 453). We, then, explored the $\chi^2$ for spins from 0 to 1.0 with stepsize of 0.01 for them (Figure \ref{fig:Fe_step}). These four models all obtain moderate spin black holes at 90\% statistical confidence level, but when the iron abundance is higher than $\sim{5.5}$, at more than 90\% statistical confidence level, we note the reduced sensitivity of models to large values of spin. The increase of iron abundance induces more photoelectric absorption making the Fe K-edge near 8 keV deeper. At the same time, the strength of the Fe k emission in the band of 6-8 keV increases \citep{gar2013}. More fluorescent iron photons near the black hole would be scattered down below $\sim{6}$ keV to make a stronger red wing, as expected.

%==============================================================

%figre6
%==============================================================
%figre7
%==============================================================

\begin{figure}
\centering
	\includegraphics[scale=0.51,angle=0]{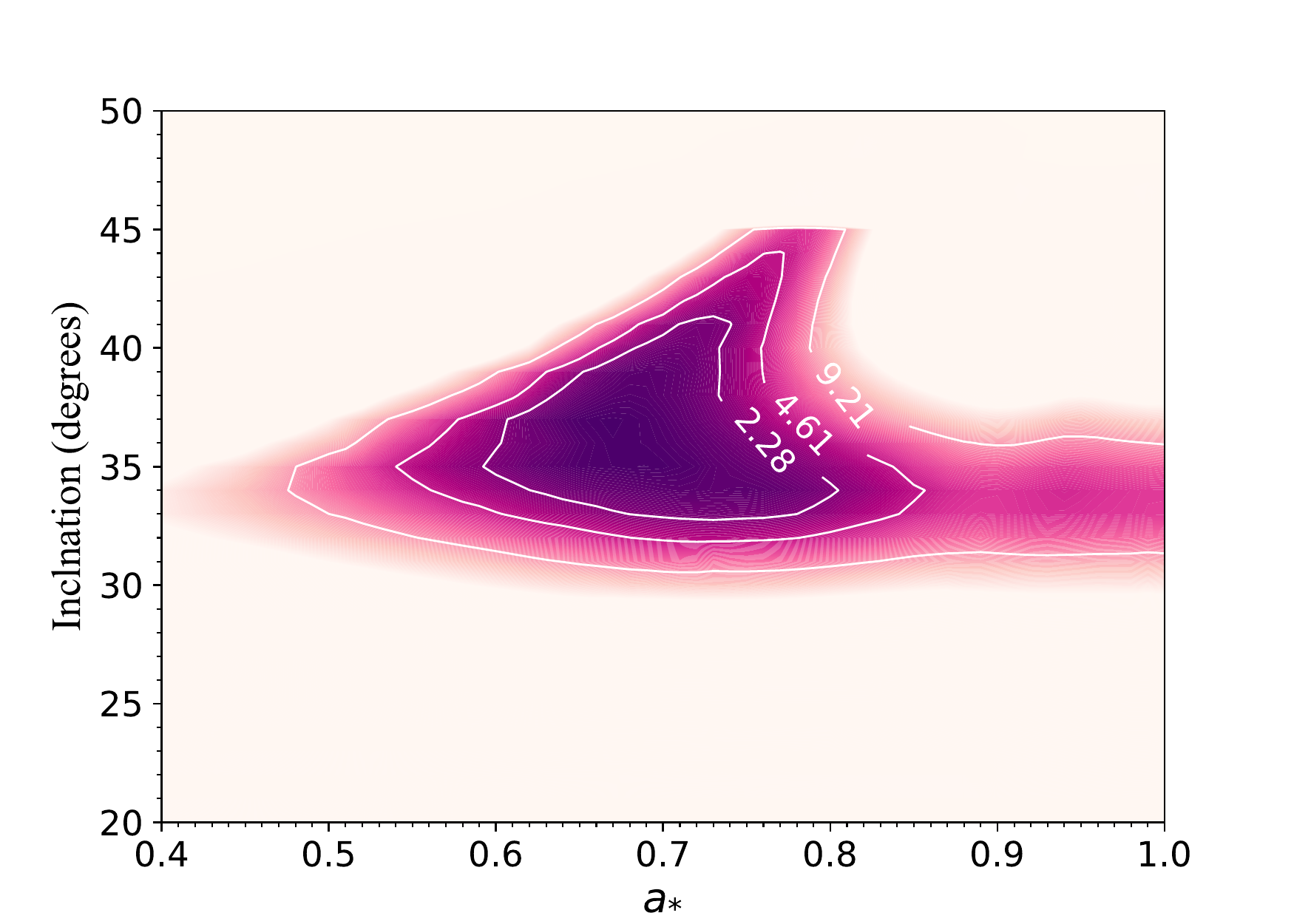}
    \caption{Joint-fit to Spec. A-G using model {\tt crabcor*TBabs*(diskbb+relxill}) (Model 1).  Contours of $\chi ^2$ with 68\%, 90\% and 99\% confidence levels for the spin parameter $a_{*}$ and inclination angle $i$ are shown in the above plot. We fit spectra for 30 spaced values of $i_r$ for the range of $20^{\circ}$-$50^{\circ}$ and 60 spaced values of the $a_*$ for the range of $0.4-1.0$.}
    \label{fig:relxill_ai}
\end{figure}

Then, we further explored the dependence of the spin parameter $a_*$ on the inclination angle $i$. We fit spectra for 60 evenly-spaced values of the $a_*$ in the range of 0.4-1.0 and 30 evenly-spaced values of the $i$ with the range of 20$^{\circ}$-50$^{\circ}$ (Figure \ref{fig:relxill_ai}). When $i$ is larger than $\sim{36}^{\circ}$, a positive relationship is shown. Moreover, the model lost the ability to give an upper limit on the spin parameter at 99\% confidence level.

The inclination angle ($36.3_{-3.4}^{+5.3}$ degrees) in this paper is consistent with the value ($32_{-4}^{+3}$ degrees) in \citet{mor2014}, which may indicate that the inclination angle of the inner disc is misaligned with the orbital inclination angle ($\sim{21}^{\circ}$). However, for a transient system, the timescale for accretion to torque the black hole into alignment is approximated to be $10^{6}-10^{8}$ years \citep{mar2008}. Therefore, the alignment is expected to occur early in the typical lifetime of transients, which are characteristically Gyrs old \citep{whi1998,fra2013}. In our estimation, the most likely resolution to this apparent tension lies in the reflection modeling. The inclination angle estimation via X-ray reflection fitting method is principally determined by the blue wing of the broad Fe line. The high density model leads to increasing soft X-ray flux. Recent reflection analyses of Cyg X-1 by \citet{joh2018}, and GX 339-4 by \citet{gar2015} and \citet{jia2019} suggest that reflection models which underestimate the density of disc introduce systematic changes of order ~10$\degr$ in the inclination angle. 

%=============================================================
%figre7
%=======================================================

\citet{sha2006} first reported the spin of the black hole in 4U 1543 via the continuum-fitting method. They estimated its spin to be $0.8\pm0.1$. 
Then, \citet{mil2009} and \citet{mor2014} reported two spin measurements, $0.3\pm0.1$ and $0.43^{+0.22}_{-0.31}$, respectively, both constrained by combining the continuum-fitting and X-ray reflection fitting methods. These three works utilized the dynamical parameters which were reported in \citet{par2004}, but found conflicting values of spin. For measuring the spin of a black hole via the continuum-fitting method to succeed, one constrains the size of the emitting region via the efficient blackbody-like property of the optically thick disc. To relate the emitting area to a dimensionless ISCO and thereby spin, it is critical to have accurate measurements of the distance to the source, the mass of the black hole, and the inclination angle of the accretion disc \citep{mcc2011,mcc2006,gou2011}.

The spin and the inclination angle measured by modeling the reflection emission with {\tt relxill} in this paper is consistent with those reported by \citet{mor2014}. However, the spin measurement is in conflict with the one reported by \citet{mil2009}. \citet{mil2009} assumed the inclination angle of accretion disc is equal to the orbital inclination angle.

Accordingly, we test the implication of the lower inclination angle on the spin measurement. When the inclination angle parameter in REXILL model was fixed at the orbital inclination angle of 21.0$^{\circ}$, which is considered as Model 2, we find that the fit becomes worse than when the inclination angle is free ($\Delta \chi^2 = 53.51$ for increasing 1 d.o.f). The best-fitting values are listed in Table \ref{tab:relxill2}. The temperature and the normalization of the thermal emission do not change significantly. The photon index parameters become smaller than the Newtonian value ($q=3$), which indicates the coronal model changes from a compact geometry ($q>3$ in Model 1) to extended. As for reflected emission, the emissivity index decreased. The spin pegged at 0.998 in this condition, possibly owning to the higher iron abundance of $7.72_{-1.62}^{+1.23}$, a more nonphysical value. The ionization state becomes much higher, and the reflection fraction becomes smaller. 

\begin{figure}
    \centering
	\includegraphics[scale=0.45,angle=0]{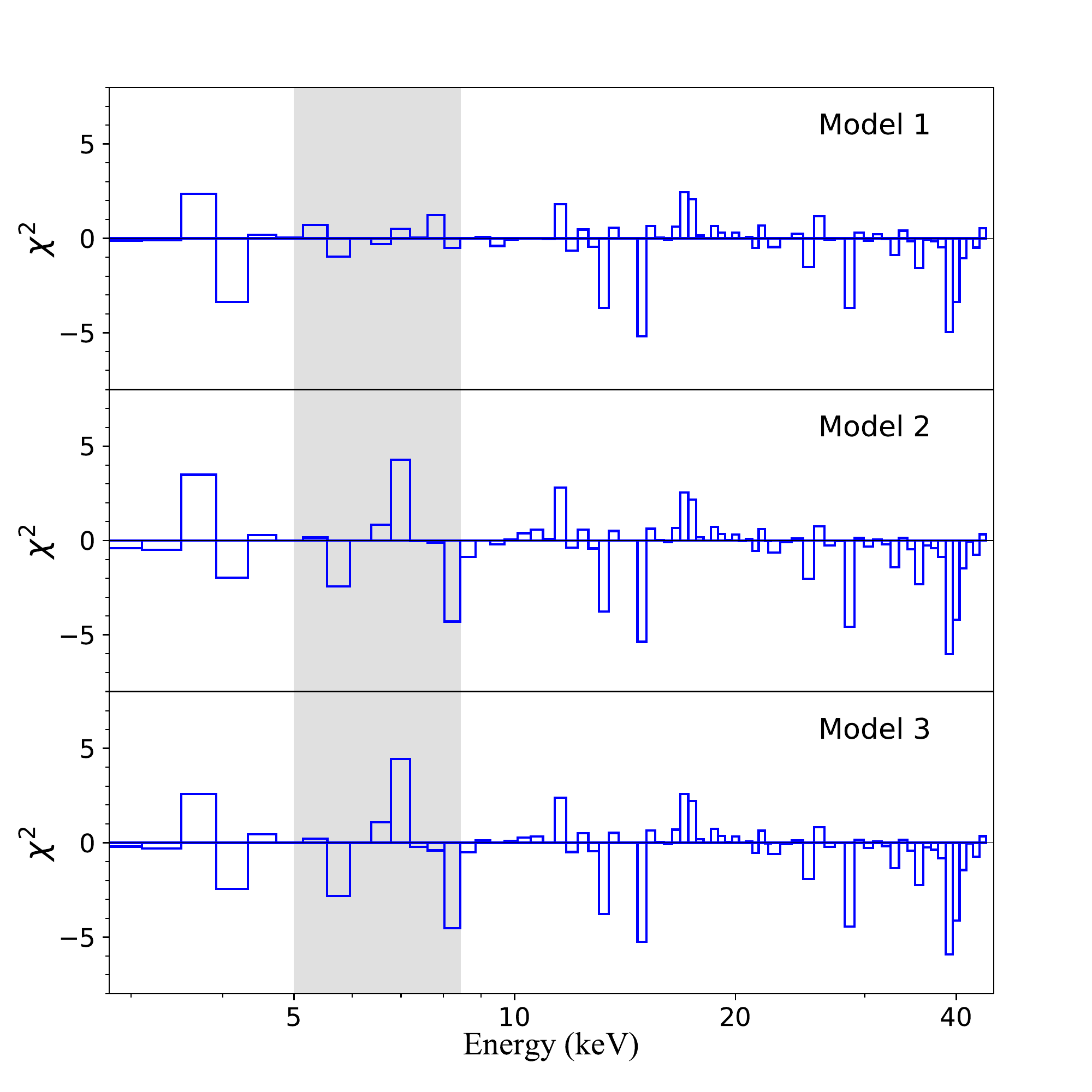}
    \caption{Contributions to $\chi^2$ for Spec. A resulting from fitting Model 1-3. From top to bottom, the d.o.f increases, starting with the inclination angle and the spin are free; the inclination angle is set at 21.0 degrees and the spin is free; the the inclination angle is set at 21.0 degrees and the spin is set at 0.3. We also included a semi-transparent box to highlight the most pronounced changes in comparing the model differences ($\sim$5-8 keV). For the other six spectra, the comparable residual plots are qualitatively similar.}
    \label{fig:comp}
\end{figure}

As an extension of Model 2, we define a new Model 3 in which we keep the inclination fixed at 21.0$^{\circ}$ and also fix the spin parameter to the value found by \citet{mil2009}: $a_{*}$=0.3. The best-fitting values are listed in Table \ref{tab:relxill3}. Compared to Model 2, its $\chi^2$ increased 12.99 for 1 d.o.f. Except the iron abundance was constrained at $4.50_{-0.21}^{+0.46}$, which is lower than that in Model 2, other parameters were not appreciably affected. We also plot the contribution to $\chi^2$ for Spec. A resulting from Model 1-3 in Figure \ref{fig:comp}. The most pronounced changes in comparing the model differences are residuals around the iron line region ($\sim$5.0-8.0 keV).
%==============================================================
%Table 3
\begin{table*}
    \caption{Best-fitting parameters for Model 2: {\tt crabcor*TBabs}*({\tt diskbb}+{\tt relxill}), in which $i$ = 21 deg}
    \begin{threeparttable}[b]
    \begin{center}
    \label{tab:relxill2}
    \footnotesize
        \begin{tabular}{llccccccc}
        \toprule
        Model & Parameter &Spec. A & Spec. B & Spec. C & Spec. D & Spec. E & Spec. F & Spec. G \\
        \midrule
        {\tt crabcor}&$C$  &\multicolumn{7}{c}{1.097  (f)}\\
        \specialrule{0em}{1pt}{1.1pt}
        &$\Delta\Gamma$  &\multicolumn{7}{c}{0.01  (f)}\\
        \specialrule{0em}{1pt}{1.1pt}
        {\tt TBabs}&$N_\mathrm{H}$ ($ cm^{-2}$) &\multicolumn{7}{c}{$4.0\times10^{21}$ (f)}\\
        \specialrule{0em}{1pt}{1.1pt}
        {\tt diskbb}   &$T_\mathrm{col}$(keV)&$0.842_{-0.013}^{+0.015}$&$0.814_{-0.015}^{+0.016}$&$0.829\pm0.015$&$0.784\pm0.013$&$0.783\pm0.007$&$0.770_{-0.007}^{+0.008}$&$0.770_{-0.011}^{+0.010}$   \\
        \specialrule{0em}{1pt}{1.1pt}
        &$N_\mathrm{DISC}$&$4349_{-291}^{+220}$&$4471_{-374}^{+376}$&$3778_{-301}^{+229}$&$5021_{-383}^{+388}$&$5124_{-216}^{+221}$&$5110_{-272}^{+251}$&$5052_{-306}^{+355}$ \\
        \specialrule{0em}{1pt}{1.1pt}
        {\tt relxill}
        &$q$&$2.72_{-0.13}^{+0.10}$&$2.69\pm0.13$&$2.44\pm0.17$&$2.78_{-0.14}^{+0.08}$&$2.67\pm0.09$&$2.59_{-0.06}^{+0.11}$&$2.76_{-0.15}^{+0.10}$\\
        \specialrule{0em}{1pt}{1.1pt}
        &$a_*$&\multicolumn{7}{c}{$> 0.83$ }\\
        \specialrule{0em}{1pt}{1.1pt}
        &$i$ (deg)&\multicolumn{7}{c}{$21.0$ (f)}\\
        \specialrule{0em}{1pt}{1.1pt}
        &$A_\mathrm{Fe}$&\multicolumn{7}{c}{$7.72_{-1.62}^{+1.23}$ }\\
        \specialrule{0em}{1pt}{1.1pt}
        &$\Gamma_\mathrm{r}$&$2.64_{-0.03}^{+0.04}$&$2.48_{-0.03}^{+0.04}$&$2.51_{-0.03}^{+0.04}$&$2.56_{-0.04}^{+0.09}$&$2.52\pm0.03$&$2.53_{-0.02}^{+0.04}$&$2.59_{-0.05}^{+0.14}$  \\
        \specialrule{0em}{1pt}{1.1pt}
        &$\mathrm{log}\xi$&$4.52_{-0.15}^{+4.52}$&$4.19_{-0.35}^{+0.25}$&$4.26_{-0.34}^{+0.15}$&$4.07_{-0.34}^{+0.25}$&$4.29_{-0.20}^{+0.09}$&$4.19_{-0.23}^{+0.16}$&$4.01_{-0.23}^{+0.35}$\\
        \specialrule{0em}{1pt}{1.1pt}
        &$R_\mathrm{r}$&$0.21_{-0.09}^{+0.06}$&$0.23_{-0.01}^{+0.06}$&$0.18_{-0.03}^{+0.04}$&$0.3_{-0.04}^{+0.06}$&$0.32\pm0.04$&$0.32_{-0.04}^{+0.05}$&$0.36_{-0.06}^{+0.13}$ \\
        \specialrule{0em}{1pt}{1.1pt}
        &$N_\mathrm{r}$&$0.111_{-0.011}^{+0.015}$&$0.063_{-0.008}^{+0.010}$&$0.064_{-0.007}^{+0.010}$&$0.04_{-0.003}^{+0.015}$&$0.032_{-0.003}^{+0.004}$&$0.023_{-0.003}^{+0.004}$&$0.031_{-0.005}^{+0.016}$ \\
        \midrule
        & $\chi^2$ &\multicolumn{7}{c}{443.71} \\
        & $\nu$ & \multicolumn{7}{c}{453} \\
        & $\chi^2_{\nu}$ & \multicolumn{7}{c}{0.98} \\
        \bottomrule
        \end{tabular}
    \begin{tablenotes}
     \item[] \textbf{Notes.} Columns 3-9 show successively the results of Spec. A-G. The parameters with ``f'' in parenthesis indicate they were fixed at values given. All errors for one parameter of interest were calculated with 90\% confidence level.
     \item \textbf{1.} Parameters including the spin $a_*$ and the iron abundance $A_\mathrm{Fe}$ of {\tt relxill} were linked together among different spectra.
     \item \textbf{2.} Parameters including the temperature $T_\mathrm{col}$, normalization constant $N_\mathrm{DISC}$, emissivity index $q$, photon index $\Gamma _\mathrm{r}$, ionization state $\mathrm{log}\xi$, reflection fraction $R_\mathrm{r}$ and the normalization $N_\mathrm{r}$ were independent for each spectrum.
    \end{tablenotes}
    \end{center}
    \end{threeparttable}
\end{table*}
%==============================================
%==============================================
%==============================================================

%Table 3
\begin{table*}
    \caption{Best-fitting parameters for Model 3: {\tt crabcor*TBabs}*({\tt diskbb}+{\tt relxill}), in which $i$ = 21 deg and $a_{*}$ = 0.30}
    \begin{threeparttable}[b]
    \begin{center}
    \label{tab:relxill3}
    \footnotesize
        \begin{tabular}{llccccccc}
        \toprule
        Model & Parameter &Spec. A & Spec. B & Spec. C & Spec. D & Spec. E & Spec. F & Spec. G \\
        \midrule
        {\tt crabcor}&$C$  &\multicolumn{7}{c}{1.097  (f)}\\
        \specialrule{0em}{1pt}{1.1pt}
        &$\Delta\Gamma$  &\multicolumn{7}{c}{0.01  (f)}\\
        \specialrule{0em}{1pt}{1.1pt}
        {\tt TBabs}&$N_\mathrm{H}$ ($ cm^{-2}$) &\multicolumn{7}{c}{$4.0\times10^{21}$(f)}\\
        \specialrule{0em}{1pt}{1.1pt}
        {\tt diskbb}    &$T_\mathrm{col}$(keV)&$0.853\pm0.013$&$0.83\pm0.015$&$0.837_{-0.016}^{+0.013}$&$0.798\pm0.009$&$0.791_{-0.008}^{+0.007}$&$0.778_{-0.007}^{+0.006}$&$0.783_{-0.010}^{+0.008}$\\
        \specialrule{0em}{1pt}{1.1pt}
        &$N_\mathrm{DISC}$&$4127_{-172}^{+262}$&$4103_{-341}^{+346}$&$3586_{-242}^{+331}$&$4595_{-259}^{+324}$&$4868_{-212}^{+250}$&$4862_{-192}^{+239}$&$4603_{-241}^{+360}$\\
        \specialrule{0em}{1pt}{1.1pt}
        {\tt relxill}
        &$q$&$2.75\pm0.15$&$2.75_{-0.15}^{+0.07}$&$2.49_{-0.20}^{+0.18}$&$2.76_{-0.19}^{+0.15}$&$2.72_{-0.11}^{+0.10}$&$2.61_{-0.13}^{+0.12}$&$2.65\pm0.23$\\
        \specialrule{0em}{1pt}{1.1pt}
        &$a_*$&\multicolumn{7}{c}{$0.30$ (f)}\\
        \specialrule{0em}{1pt}{1.1pt}
        &$i$ (deg)&\multicolumn{7}{c}{$21.0$ (f)}\\
        \specialrule{0em}{1pt}{1.1pt}
        &$A_\mathrm{Fe}$&\multicolumn{7}{c}{$4.5_{-0.21}^{+0.46}$  }\\
        \specialrule{0em}{1pt}{1.1pt}
        &$\Gamma_\mathrm{r}$&$2.65\pm0.03$&$2.5_{-0.03}^{+0.05}$&$2.54\pm0.03$&$2.62_{-0.06}^{+0.08}$&$2.54\pm0.02$&$2.56\pm0.03$&$2.73_{-0.12}^{+0.06}$ \\
        \specialrule{0em}{1pt}{1.1pt}
        &$\mathrm{log}\xi$&$4.31_{-0.63}^{+0.22}$&$3.89_{-0.36}^{+0.29}$&$4.03_{-0.27}^{+0.30}$&$3.73_{-0.21}^{+0.31}$&$4.06_{-0.18}^{+0.24}$&$4.0_{-0.20}^{+0.17}$&$3.63_{-0.16}^{+0.27}$\\
        \specialrule{0em}{1pt}{1.1pt}
        &$R_\mathrm{r}$&$0.18\pm0.03$&$0.19_{-0.02}^{+0.03}$&$0.17\pm0.02$&$0.29\pm0.06$&$0.27_{-0.02}^{+0.04}$&$0.29\pm0.03$&$0.39_{-0.06}^{+0.09}$  \\
        \specialrule{0em}{1pt}{1.1pt}
        &$N_\mathrm{r}$&$0.115_{-0.012}^{+0.025}$&$0.069_{-0.008}^{+0.012}$&$0.07_{-0.008}^{+0.009}$&$0.049_{-0.009}^{+0.013}$&$0.034\pm0.003$&$0.026\pm0.003$&$0.046_{-0.014}^{+0.009}$ \\
        \midrule
        & $\chi^2$ &\multicolumn{7}{c}{456.69} \\
        & $\nu$ & \multicolumn{7}{c}{454} \\
        & $\chi^2_{\nu}$ & \multicolumn{7}{c}{1.01} \\
        \bottomrule
        \end{tabular}
    \begin{tablenotes}
     \item \textbf{Notes.} Columns 3-9 show successively the results of Spec. A-G. The parameters with ``f'' in parenthesis indicate they were fixed at values given. All errors for one parameter of interest were calculated with 90\% confidence level.
     \item \textbf{1.} The iron abundance $A_\mathrm{Fe}$ of {\tt relxill} were linked together among different spectra.
     \item \textbf{2.} Parameters including the temperature $T_\mathrm{col}$, normalization constant $N_\mathrm{DISC}$, emissivity index $q$, photon index $\Gamma _\mathrm{r}$, ionization state $\mathrm{log}\xi$, reflection fraction $R_\mathrm{r}$ and the normalization $N_\mathrm{r}$ were independent for each spectrum.
    \end{tablenotes}
    \end{center}
    \end{threeparttable}
\end{table*}

%==============================================================
\section{Conclusions}
\label{sec:con}
We have measured the spin of 4U 1543 via modeling its reflected components in 7 SPL state observations carefully. The spectra consist of 4 different components: the galactic absorption, thermal emission from the accretion disc, power-law emission and reflected emission. We did joint-fit to all the spectra to improve the signal-to-noise ratio of X-ray reflected component. We used the reflection model, {\tt relxill}, to fit the data. We find a super-solar iron abundance for the disc. At the same time, the disc is highly ionized. 

The model with free inclination angle and spin (Model 1) describes the spectra best. The inclination angle of the inner accretion disc is constrained to be $36.3_{-3.4}^{+5.3}$ degrees at 90\% statistical confidence. When the inclination angle is fixed at the orbital inclination value of 21.0$^{\circ}$ in Model 2 or Model 3, the statistic becomes significantly worse, and the spin is larger than 0.83. The best-fitting inclination differs from that of the orbital plane by more than 10 degrees. This may be owed to the systematic limitations of current models which underestimate the density of disc.

Our results indicate a moderate rotation rate for the black hole in 4U 1543. The spin parameter is established to be $0.67_{-0.08}^{+0.15}$ at 90\% statistical confidence. At the 99\% statistical confidence level, we exclude spins below $a_{*}$< 0.5 (which also excludes any retrograde geometries).
  
%==============================================================
\section*{Acknowledgements}

We thank the useful discussion with Prof. J.Orosz, Prof. Youjun Lu, Dr. Erlin Qiao, Dr. Weiwei Xu, and Dr. Zhu Liu. We would also like to thank the reviewer for his/her valuable input. Lijun Gou are supported by the National Program on Key Research and Development Project through grant No. 2016YFA0400804, and by the National Natural Science Foundation of China with grant No. U1838114, and by the Strategic Priority Research Program of the Chinese Academy of Sciences through grant No. XDB23040100. We also thank RXTE/PCA public data and facilities. This work is made under the help with tools available on Astrophysics Science Archive Research Centre (HEASARC), belonging to NASA's Coddard Space Flight Centre (GSFC). 
%%%%%%%%%%%%%%%%%%%%%%%%%%%%%%%%%%%%%%%%%%%%%%%%%%



\bibliographystyle{mnras}
\bibliography{YTNotes} % if your bibtex file is called example.bib

\begin{thebibliography}{}
\makeatletter
\relax
\def\mn@urlcharsother{\let\do\@makeother \do\$\do\&\do\#\do\^\do\_\do\%\do\~}
\def\mn@doi{\begingroup\mn@urlcharsother \@ifnextchar [ {\mn@doi@}
  {\mn@doi@[]}}
\def\mn@doi@[#1]#2{\def\@tempa{#1}\ifx\@tempa\@empty \href
  {http://dx.doi.org/#2} {doi:#2}\else \href {http://dx.doi.org/#2} {#1}\fi
  \endgroup}
\def\mn@eprint#1#2{\mn@eprint@#1:#2::\@nil}
\def\mn@eprint@arXiv#1{\href {http://arxiv.org/abs/#1} {{\tt arXiv:#1}}}
\def\mn@eprint@dblp#1{\href {http://dblp.uni-trier.de/rec/bibtex/#1.xml}
  {dblp:#1}}
\def\mn@eprint@#1:#2:#3:#4\@nil{\def\@tempa {#1}\def\@tempb {#2}\def\@tempc
  {#3}\ifx \@tempc \@empty \let \@tempc \@tempb \let \@tempb \@tempa \fi \ifx
  \@tempb \@empty \def\@tempb {arXiv}\fi \@ifundefined
  {mn@eprint@\@tempb}{\@tempb:\@tempc}{\expandafter \expandafter \csname
  mn@eprint@\@tempb\endcsname \expandafter{\@tempc}}}

\bibitem[\protect\citeauthoryear{{Arnaud}}{{Arnaud}}{1996}]{arn1996}
{Arnaud} K.~A.,  1996, in {Jacoby} G.~H.,  {Barnes} J.,  eds,  ASP Conf. Ser.
  Vol. 101, Astronomical Data Analysis Software and Systems V. p.~17

\bibitem[\protect\citeauthoryear{{Brown} \& {Bethe}}{{Brown} \&
  {Bethe}}{1994}]{bro1994}
{Brown} G.~E.,  {Bethe} H.~A.,  1994, \mn@doi [\apj] {10.1086/173844}, \href
  {https://ui.adsabs.harvard.edu/abs/1994ApJ...423..659B} {423, 659}

\bibitem[\protect\citeauthoryear{{Chen}, {Gou}, {McClintock}, {Steiner}, {Wu},
  {Xu}, {Orosz}  \& {Xiang}}{{Chen} et~al.}{2016}]{che2016}
{Chen} Z.,  {Gou} L.,  {McClintock} J.~E.,  {Steiner} J.~F.,  {Wu} J.,  {Xu}
  W.,  {Orosz} J.~A.,   {Xiang} Y.,  2016, \mn@doi [\apj]
  {10.3847/0004-637X/825/1/45}, \href
  {http://ads.bao.ac.cn/abs/2016ApJ...825...45C} {825, 45}

\bibitem[\protect\citeauthoryear{{Dauser}, {Wilms}, {Reynolds}  \&
  {Brenneman}}{{Dauser} et~al.}{2010}]{dau2010}
{Dauser} T.,  {Wilms} J.,  {Reynolds} C.~S.,   {Brenneman} L.~W.,  2010,
  \mn@doi [\mnras] {10.1111/j.1365-2966.2010.17393.x}, \href
  {http://ads.bao.ac.cn/abs/2010MNRAS.409.1534D} {409, 1534}

\bibitem[\protect\citeauthoryear{{Dauser} et~al.,}{{Dauser}
  et~al.}{2012}]{dau2012}
{Dauser} T.,  et~al., 2012, \mn@doi [\mnras]
  {10.1111/j.1365-2966.2011.20356.x}, \href
  {http://adsabs.harvard.edu/abs/2012MNRAS.422.1914D} {422, 1914}

\bibitem[\protect\citeauthoryear{{Dauser}, {Garcia}, {Wilms}, {B{\"o}ck},
  {Brenneman}, {Falanga}, {Fukumura}  \& {Reynolds}}{{Dauser}
  et~al.}{2013}]{dau2013}
{Dauser} T.,  {Garcia} J.,  {Wilms} J.,  {B{\"o}ck} M.,  {Brenneman} L.~W.,
  {Falanga} M.,  {Fukumura} K.,   {Reynolds} C.~S.,  2013, \mn@doi [\mnras]
  {10.1093/mnras/sts710}, \href
  {http://adsabs.harvard.edu/abs/2013MNRAS.430.1694D} {430, 1694}

\bibitem[\protect\citeauthoryear{{Dauser}, {Garc{\'{\i}}a}, {Parker}, {Fabian}
  \& {Wilms}}{{Dauser} et~al.}{2014}]{dau2014}
{Dauser} T.,  {Garc{\'{\i}}a} J.,  {Parker} M.~L.,  {Fabian} A.~C.,   {Wilms}
  J.,  2014, \mn@doi [\mnras] {10.1093/mnrasl/slu125}, \href
  {http://ads.bao.ac.cn/abs/2014MNRAS.444L.100D} {444, L100}

\bibitem[\protect\citeauthoryear{{Ebisawa} et~al.,}{{Ebisawa}
  et~al.}{1994}]{ebi1994}
{Ebisawa} K.,  et~al., 1994, \pasj, \href
  {http://ads.bao.ac.cn/abs/1994PASJ...46..375E} {46, 375}

\bibitem[\protect\citeauthoryear{{Fragos} et~al.,}{{Fragos}
  et~al.}{2013}]{fra2013}
{Fragos} T.,  et~al., 2013, \mn@doi [\apj] {10.1088/0004-637X/764/1/41}, \href
  {https://ui.adsabs.harvard.edu/abs/2013ApJ...764...41F} {764, 41}

\bibitem[\protect\citeauthoryear{{Garc{\'{\i}}a} \& {Kallman}}{{Garc{\'{\i}}a}
  \& {Kallman}}{2010}]{gar2010}
{Garc{\'{\i}}a} J.,  {Kallman} T.~R.,  2010, \mn@doi [\apj]
  {10.1088/0004-637X/718/2/695}, \href
  {http://adsabs.harvard.edu/abs/2010ApJ...718..695G} {718, 695}

\bibitem[\protect\citeauthoryear{{Garc{\'{\i}}a}, {Kallman}  \&
  {Mushotzky}}{{Garc{\'{\i}}a} et~al.}{2011}]{gar2011}
{Garc{\'{\i}}a} J.,  {Kallman} T.~R.,   {Mushotzky} R.~F.,  2011, \mn@doi
  [\apj] {10.1088/0004-637X/731/2/131}, \href
  {http://adsabs.harvard.edu/abs/2011ApJ...731..131G} {731, 131}

\bibitem[\protect\citeauthoryear{{Garc{\'{\i}}a}, {Dauser}, {Reynolds},
  {Kallman}, {McClintock}, {Wilms}  \& {Eikmann}}{{Garc{\'{\i}}a}
  et~al.}{2013}]{gar2013}
{Garc{\'{\i}}a} J.,  {Dauser} T.,  {Reynolds} C.~S.,  {Kallman} T.~R.,
  {McClintock} J.~E.,  {Wilms} J.,   {Eikmann} W.,  2013, \mn@doi [\apj]
  {10.1088/0004-637X/768/2/146}, \href
  {http://ads.bao.ac.cn/abs/2013ApJ...768..146G} {768, 146}

\bibitem[\protect\citeauthoryear{{Garc{\'{\i}}a} et~al.,}{{Garc{\'{\i}}a}
  et~al.}{2014a}]{gar2014a}
{Garc{\'{\i}}a} J.,  et~al., 2014a, \mn@doi [\apj]
  {10.1088/0004-637X/782/2/76}, \href
  {http://ads.bao.ac.cn/abs/2014ApJ...782...76G} {782, 76}

\bibitem[\protect\citeauthoryear{{Garc{\'{\i}}a}, {McClintock}, {Steiner},
  {Remillard}  \& {Grinberg}}{{Garc{\'{\i}}a} et~al.}{2014b}]{gar2014b}
{Garc{\'{\i}}a} J.~A.,  {McClintock} J.~E.,  {Steiner} J.~F.,  {Remillard}
  R.~A.,   {Grinberg} V.,  2014b, \mn@doi [\apj] {10.1088/0004-637X/794/1/73},
  \href {http://adsabs.harvard.edu/abs/2014ApJ...794...73G} {794, 73}

\bibitem[\protect\citeauthoryear{{Garc{\'{\i}}a}, {Steiner}, {McClintock},
  {Remillard}, {Grinberg}  \& {Dauser}}{{Garc{\'{\i}}a} et~al.}{2015}]{gar2015}
{Garc{\'{\i}}a} J.~A.,  {Steiner} J.~F.,  {McClintock} J.~E.,  {Remillard}
  R.~A.,  {Grinberg} V.,   {Dauser} T.,  2015, \mn@doi [\apj]
  {10.1088/0004-637X/813/2/84}, \href
  {http://ads.bao.ac.cn/abs/2015ApJ...813...84G} {813, 84}

\bibitem[\protect\citeauthoryear{{Garc{\'{\i}}a} et~al.,}{{Garc{\'{\i}}a}
  et~al.}{2018}]{Gar2018}
{Garc{\'{\i}}a} J.~A.,  et~al., 2018, \mn@doi [\apj]
  {10.3847/1538-4357/aad231}, \href
  {http://ads.bao.ac.cn/abs/2018ApJ...864...25G} {864, 25}

\bibitem[\protect\citeauthoryear{{Gou} et~al.,}{{Gou} et~al.}{2009}]{gou2009}
{Gou} L.,  et~al., 2009, \mn@doi [\apj] {10.1088/0004-637X/701/2/1076}, \href
  {http://ads.bao.ac.cn/abs/2009ApJ...701.1076G} {701, 1076}

\bibitem[\protect\citeauthoryear{{Gou} et~al.,}{{Gou} et~al.}{2011}]{gou2011}
{Gou} L.,  et~al., 2011, \mn@doi [\apj] {10.1088/0004-637X/742/2/85}, \href
  {http://ads.bao.ac.cn/abs/2011ApJ...742...85G} {742, 85}

\bibitem[\protect\citeauthoryear{{Harmon}, {Wilson}, {Finger}, {Paciesas},
  {Rubin}  \& {Fishman}}{{Harmon} et~al.}{1992}]{har1992}
{Harmon} B.~A.,  {Wilson} R.~B.,  {Finger} M.~H.,  {Paciesas} W.~S.,  {Rubin}
  B.~C.,   {Fishman} G.~J.,  1992, \iaucirc, \href
  {http://adsabs.harvard.edu/abs/1992IAUC.5504....1H} {5504}

\bibitem[\protect\citeauthoryear{{Iwasawa} et~al.,}{{Iwasawa}
  et~al.}{1997}]{iwa1997}
{Iwasawa} K.,  et~al., 1997, in {Makino} F.,  {Mitsuda} K.,  eds, X-Ray Imaging
  and Spectroscopy of Cosmic Hot Plasmas. p.~247

\bibitem[\protect\citeauthoryear{{Jahoda}, {Swank}, {Giles}, {Stark},
  {Strohmayer}, {Zhang}  \& {Morgan}}{{Jahoda} et~al.}{1996}]{jah1996}
{Jahoda} K.,  {Swank} J.~H.,  {Giles} A.~B.,  {Stark} M.~J.,  {Strohmayer} T.,
  {Zhang} W.,   {Morgan} E.~H.,  1996, in {Siegmund} O.~H.,  {Gummin} M.~A.,
  eds,  Society of Photo-Optical Instrumentation Engineers (SPIE) Conference
  Series Vol. 2808, \procspie. pp 59--70, \mn@doi{10.1117/12.256034}

\bibitem[\protect\citeauthoryear{{Jahoda}, {Markwardt}, {Radeva}, {Rots},
  {Stark}, {Swank}, {Strohmayer}  \& {Zhang}}{{Jahoda} et~al.}{2006}]{jah2006}
{Jahoda} K.,  {Markwardt} C.~B.,  {Radeva} Y.,  {Rots} A.~H.,  {Stark} M.~J.,
  {Swank} J.~H.,  {Strohmayer} T.~E.,   {Zhang} W.,  2006, \mn@doi [\apjs]
  {10.1086/500659}, \href
  {https://ui.adsabs.harvard.edu/abs/2006ApJS..163..401J} {163, 401}

\bibitem[\protect\citeauthoryear{{Jiang}, {Fabian}, {Wang}, {Walton},
  {Garc{\'\i}a}, {Parker}, {Steiner}  \& {Tomsick}}{{Jiang}
  et~al.}{2019}]{jia2019}
{Jiang} J.,  {Fabian} A.~C.,  {Wang} J.,  {Walton} D.~J.,  {Garc{\'\i}a} J.~A.,
   {Parker} M.~L.,  {Steiner} J.~F.,   {Tomsick} J.~A.,  2019, \mn@doi [\mnras]
  {10.1093/mnras/stz095}, \href
  {https://ui.adsabs.harvard.edu/abs/2019MNRAS.484.1972J} {484, 1972}

\bibitem[\protect\citeauthoryear{{Kerr}}{{Kerr}}{1963}]{ker1963}
{Kerr} R.~P.,  1963, \mn@doi [Phys. Rev. Lett.] {10.1103/PhysRevLett.11.237},
  \href {http://adsabs.harvard.edu/abs/1963PhRvL..11..237K} {11, 237}

\bibitem[\protect\citeauthoryear{{Kitamoto}, {Miyamoto}, {Tsunemi}, {Makishima}
   \& {Nakagawa}}{{Kitamoto} et~al.}{1984}]{kit1984}
{Kitamoto} S.,  {Miyamoto} S.,  {Tsunemi} H.,  {Makishima} K.,   {Nakagawa} M.,
   1984, \pasj, \href {http://adsabs.harvard.edu/abs/1984PASJ...36..799K} {36,
  799}

\bibitem[\protect\citeauthoryear{Li, Zimmerman, Narayan  \& McClintock}{Li
  et~al.}{2005}]{Li2005}
Li L.-X.,  Zimmerman E.~R.,  Narayan R.,   McClintock J.~E.,  2005, \mn@doi
  [\apjs] {10.1086/428089}, 157, 335

\bibitem[\protect\citeauthoryear{{Martin}, {Tout}  \& {Pringle}}{{Martin}
  et~al.}{2008}]{mar2008}
{Martin} R.~G.,  {Tout} C.~A.,   {Pringle} J.~E.,  2008, \mn@doi [\mnras]
  {10.1111/j.1365-2966.2008.13148.x}, \href
  {https://ui.adsabs.harvard.edu/abs/2008MNRAS.387..188M} {387, 188}

\bibitem[\protect\citeauthoryear{{Matilsky}, {Giacconi}, {Gursky}, {Kellogg}
  \& {Tananbaum}}{{Matilsky} et~al.}{1972}]{mat1972}
{Matilsky} T.~A.,  {Giacconi} R.,  {Gursky} H.,  {Kellogg} E.~M.,   {Tananbaum}
  H.~D.,  1972, \mn@doi [\apjl] {10.1086/180947}, \href
  {http://adsabs.harvard.edu/abs/1972ApJ...174L..53M} {174, L53}

\bibitem[\protect\citeauthoryear{{McClintock}, {Shafee}, {Narayan},
  {Remillard}, {Davis}  \& {Li}}{{McClintock} et~al.}{2006}]{mcc2006}
{McClintock} J.~E.,  {Shafee} R.,  {Narayan} R.,  {Remillard} R.~A.,  {Davis}
  S.~W.,   {Li} L.-X.,  2006, \mn@doi [\apj] {10.1086/508457}, \href
  {http://ads.bao.ac.cn/abs/2006ApJ...652..518M} {652, 518}

\bibitem[\protect\citeauthoryear{McClintock et~al.,}{McClintock
  et~al.}{2011}]{mcc2011}
McClintock J.~E.,  et~al., 2011, \mn@doi [Class. Quantum Grav.]
  {10.1088/0264-9381/28/11/114009}, 28, 114009

\bibitem[\protect\citeauthoryear{{Miller} et~al.,}{{Miller}
  et~al.}{2002}]{mil2002}
{Miller} J.~M.,  et~al., 2002, \mn@doi [\apjl] {10.1086/341099}, \href
  {https://ui.adsabs.harvard.edu/abs/2002ApJ...570L..69M} {570, L69}

\bibitem[\protect\citeauthoryear{{Miller}, {Fabian}  \& {Lewin}}{{Miller}
  et~al.}{2003}]{mil2003}
{Miller} J.~M.,  {Fabian} A.~C.,   {Lewin} W.~H.~G.,  2003, ATel, \href
  {https://ui.adsabs.harvard.edu/abs/2003ATel..212....1M} {212, 1}

\bibitem[\protect\citeauthoryear{{Miller}, {Reynolds}, {Fabian}, {Miniutti}  \&
  {Gallo}}{{Miller} et~al.}{2009}]{mil2009}
{Miller} J.~M.,  {Reynolds} C.~S.,  {Fabian} A.~C.,  {Miniutti} G.,   {Gallo}
  L.~C.,  2009, \mn@doi [\apj] {10.1088/0004-637X/697/1/900}, \href
  {http://adsabs.harvard.edu/abs/2009ApJ...697..900M} {697, 900}

\bibitem[\protect\citeauthoryear{{Morningstar} \& {Miller}}{{Morningstar} \&
  {Miller}}{2014}]{mor2014}
{Morningstar} W.~R.,  {Miller} J.~M.,  2014, \mn@doi [\apjl]
  {10.1088/2041-8205/793/2/L33}, \href
  {http://adsabs.harvard.edu/abs/2014ApJ...793L..33M} {793, L33}

\bibitem[\protect\citeauthoryear{Nayakshin \& Kallman}{Nayakshin \&
  Kallman}{2001}]{nay2001}
Nayakshin S.,  Kallman T.~R.,  2001, \mn@doi [\apj] {10.1086/318250}, 546, 406

\bibitem[\protect\citeauthoryear{{Orosz}, {Jain}, {Bailyn}, {McClintock}  \&
  {Remillard}}{{Orosz} et~al.}{1998}]{oro1998}
{Orosz} J.~A.,  {Jain} R.~K.,  {Bailyn} C.~D.,  {McClintock} J.~E.,
  {Remillard} R.~A.,  1998, \mn@doi [\apj] {10.1086/305620}, \href
  {http://ads.bao.ac.cn/abs/1998ApJ...499..375O} {499, 375}

\bibitem[\protect\citeauthoryear{{Park} et~al.,}{{Park} et~al.}{2004}]{par2004}
{Park} S.~Q.,  et~al., 2004, \mn@doi [\apj] {10.1086/421511}, \href
  {http://adsabs.harvard.edu/abs/2004ApJ...610..378P} {610, 378}

\bibitem[\protect\citeauthoryear{{Parker} et~al.,}{{Parker}
  et~al.}{2015}]{par2015}
{Parker} M.~L.,  et~al., 2015, \mn@doi [\apj] {10.1088/0004-637X/808/1/9},
  \href {http://ads.bao.ac.cn/abs/2015ApJ...808....9P} {808, 9}

\bibitem[\protect\citeauthoryear{{Parker} et~al.,}{{Parker}
  et~al.}{2016}]{par2016}
{Parker} M.~L.,  et~al., 2016, \mn@doi [\apjl] {10.3847/2041-8205/821/1/L6},
  \href {https://ui.adsabs.harvard.edu/abs/2016ApJ...821L...6P} {821, L6}

\bibitem[\protect\citeauthoryear{{Remillard} \& {McClintock}}{{Remillard} \&
  {McClintock}}{2006}]{rem2006}
{Remillard} R.~A.,  {McClintock} J.~E.,  2006, \mn@doi [\araa]
  {10.1146/annurev.astro.44.051905.092532}, \href
  {http://adsabs.harvard.edu/abs/2006ARA%26A..44...49R} {44, 49}

\bibitem[\protect\citeauthoryear{{Reynolds}}{{Reynolds}}{2019}]{rey2019}
{Reynolds} C.~S.,  2019, \mn@doi [Nature Astron.] {10.1038/s41550-018-0665-z},
  \href {https://ui.adsabs.harvard.edu/abs/2019NatAs...3...41R} {3, 41}

\bibitem[\protect\citeauthoryear{{Rhoades} \& {Ruffini}}{{Rhoades} \&
  {Ruffini}}{1974}]{rho1974}
{Rhoades} C.~E.,  {Ruffini} R.,  1974, \mn@doi [Phys. Rev. Lett.]
  {10.1103/PhysRevLett.32.324}, \href
  {http://ads.bao.ac.cn/abs/1974PhRvL..32..324R} {32, 324}

\bibitem[\protect\citeauthoryear{{Shafee}, {McClintock}, {Narayan}, {Davis},
  {Li}  \& {Remillard}}{{Shafee} et~al.}{2006}]{sha2006}
{Shafee} R.,  {McClintock} J.~E.,  {Narayan} R.,  {Davis} S.~W.,  {Li} L.-X.,
  {Remillard} R.~A.,  2006, \mn@doi [\apjl] {10.1086/498938}, \href
  {http://adsabs.harvard.edu/abs/2006ApJ...636L.113S} {636, L113}

\bibitem[\protect\citeauthoryear{{Shaposhnikov}, {Jahoda}, {Markwardt}, {Swank}
   \& {Strohmayer}}{{Shaposhnikov} et~al.}{2012}]{sha2012}
{Shaposhnikov} N.,  {Jahoda} K.,  {Markwardt} C.,  {Swank} J.,   {Strohmayer}
  T.,  2012, \mn@doi [\apj] {10.1088/0004-637X/757/2/159}, \href
  {https://ui.adsabs.harvard.edu/abs/2012ApJ...757..159S} {757, 159}

\bibitem[\protect\citeauthoryear{{Steiner}, {McClintock}, {Remillard}, {Gou},
  {Yamada}  \& {Narayan}}{{Steiner} et~al.}{2010}]{ste2010}
{Steiner} J.~F.,  {McClintock} J.~E.,  {Remillard} R.~A.,  {Gou} L.,  {Yamada}
  S.,   {Narayan} R.,  2010, \mn@doi [\apjl] {10.1088/2041-8205/718/2/L117},
  \href {http://ads.bao.ac.cn/abs/2010ApJ...718L.117S} {718, L117}

\bibitem[\protect\citeauthoryear{{Steiner} et~al.,}{{Steiner}
  et~al.}{2011}]{ste2011}
{Steiner} J.~F.,  et~al., 2011, \mn@doi [\mnras]
  {10.1111/j.1365-2966.2011.19089.x}, \href
  {http://ads.bao.ac.cn/abs/2011MNRAS.416..941S} {416, 941}

\bibitem[\protect\citeauthoryear{{Timmes}, {Woosley}  \& {Weaver}}{{Timmes}
  et~al.}{1996}]{tim1996}
{Timmes} F.~X.,  {Woosley} S.~E.,   {Weaver} T.~A.,  1996, \mn@doi [\apj]
  {10.1086/176778}, \href
  {https://ui.adsabs.harvard.edu/abs/1996ApJ...457..834T} {457, 834}

\bibitem[\protect\citeauthoryear{Tomsick et~al.,}{Tomsick
  et~al.}{2018}]{joh2018}
Tomsick J.~A.,  et~al., 2018, \apj, 855, 3

\bibitem[\protect\citeauthoryear{{Toor} \& {Seward}}{{Toor} \&
  {Seward}}{1974}]{too1974}
{Toor} A.,  {Seward} F.~D.,  1974, \mn@doi [\aj] {10.1086/111643}, \href
  {https://ui.adsabs.harvard.edu/abs/1974AJ.....79..995T} {79, 995}

\bibitem[\protect\citeauthoryear{{Verner}, {Ferland}, {Korista}  \&
  {Yakovlev}}{{Verner} et~al.}{1996}]{ver1996}
{Verner} D.~A.,  {Ferland} G.~J.,  {Korista} K.~T.,   {Yakovlev} D.~G.,  1996,
  \mn@doi [\apj] {10.1086/177435}, \href
  {http://adsabs.harvard.edu/abs/1996ApJ...465..487V} {465, 487}

\bibitem[\protect\citeauthoryear{{Walton} et~al.,}{{Walton}
  et~al.}{2016}]{wal2016}
{Walton} D.~J.,  et~al., 2016, \mn@doi [\apj] {10.3847/0004-637X/826/1/87},
  \href {http://ads.bao.ac.cn/abs/2016ApJ...826...87W} {826, 87}

\bibitem[\protect\citeauthoryear{{Walton} et~al.,}{{Walton}
  et~al.}{2017}]{wal2017}
{Walton} D.~J.,  et~al., 2017, \mn@doi [\apj] {10.3847/1538-4357/aa67e8}, \href
  {https://ui.adsabs.harvard.edu/abs/2017ApJ...839..110W} {839, 110}

\bibitem[\protect\citeauthoryear{{Walton} et~al.,}{{Walton}
  et~al.}{2019}]{wal2019}
{Walton} D.~J.,  et~al., 2019, \mn@doi [\mnras] {10.1093/mnras/stz115}, \href
  {https://ui.adsabs.harvard.edu/abs/2019MNRAS.484.2544W} {484, 2544}

\bibitem[\protect\citeauthoryear{{Wang-Ji} et~al.,}{{Wang-Ji}
  et~al.}{2018}]{wan2018}
{Wang-Ji} J.,  et~al., 2018, \mn@doi [\apj] {10.3847/1538-4357/aaa974}, \href
  {http://ads.bao.ac.cn/abs/2018ApJ...855...61W} {855, 61}

\bibitem[\protect\citeauthoryear{{Wang}, {Ghasemi-Nodehi}, {Guainazzi}  \&
  {Bambi}}{{Wang} et~al.}{2017}]{wan2017}
{Wang} Y.,  {Ghasemi-Nodehi} M.,  {Guainazzi} M.,   {Bambi} C.,  2017, arXiv
  e-prints, \href {https://ui.adsabs.harvard.edu/abs/2017arXiv170307182W} {p.
  arXiv:1703.07182}

\bibitem[\protect\citeauthoryear{{White} \& {Ghosh}}{{White} \&
  {Ghosh}}{1998}]{whi1998}
{White} N.~E.,  {Ghosh} P.,  1998, \mn@doi [\apjl] {10.1086/311568}, \href
  {https://ui.adsabs.harvard.edu/abs/1998ApJ...504L..31W} {504, L31}

\bibitem[\protect\citeauthoryear{{Wilms}, {Allen}  \& {McCray}}{{Wilms}
  et~al.}{2000}]{wil2000}
{Wilms} J.,  {Allen} A.,   {McCray} R.,  2000, \mn@doi [\apj] {10.1086/317016},
  \href {http://adsabs.harvard.edu/abs/2000ApJ...542..914W} {542, 914}

\bibitem[\protect\citeauthoryear{{Xu} et~al.,}{{Xu} et~al.}{2018}]{xu2018}
{Xu} Y.,  et~al., 2018, \mn@doi [\apjl] {10.3847/2041-8213/aaa4b2}, \href
  {http://ads.bao.ac.cn/abs/2018ApJ...852L..34X} {852, L34}

\bibitem[\protect\citeauthoryear{{Zhang}, {Giles}, {Jahoda}, {Soong}, {Swank}
  \& {Morgan}}{{Zhang} et~al.}{1993}]{zha1993}
{Zhang} W.,  {Giles} A.~B.,  {Jahoda} K.,  {Soong} Y.,  {Swank} J.~H.,
  {Morgan} E.~H.,  1993, in {Siegmund} O.~H.,  ed.,  Society of Photo-Optical
  Instrumentation Engineers (SPIE) Conference Series Vol. 2006, \procspie. pp
  324--333, \mn@doi{10.1117/12.162845}

\bibitem[\protect\citeauthoryear{{Zhang}, {Cui}  \& {Chen}}{{Zhang}
  et~al.}{1997}]{zha1997}
{Zhang} S.~N.,  {Cui} W.,   {Chen} W.,  1997, \mn@doi [\apjl] {10.1086/310705},
  \href {http://adsabs.harvard.edu/abs/1997ApJ...482L.155Z} {482, L155}

\makeatother
\end{thebibliography}

%%%%%%%%%%%%%%%%%%%%%%%%%%%%%%%%%%%%%%%%%%%%%%%%%%

% Don't change these lines
\bsp	% typesetting comment
\label{lastpage}
\end{document}